\documentclass[aps,floats,showpacs,showkeys,preprintnumbers,nofootinbib]{revtex4}
\pdfoutput=1
\usepackage{amsmath,amssymb}
\usepackage[pdftex,colorlinks=false,urlcolor=blue,linkcolor=blue]{hyperref}

\usepackage{epsfig}

\usepackage{array}
\newcommand{\PreserveBackslash}[1]{\let\temp=\\#1\let\\=\temp}
\newcolumntype{C}[1]{>{\PreserveBackslash\centering}p{#1}}
\newcolumntype{R}[1]{>{\PreserveBackslash\raggedleft}p{#1}}
\newcolumntype{L}[1]{>{\PreserveBackslash\raggedright}p{#1}}

\usepackage{graphicx}
\usepackage{color}

\def\ie{{\it i.e.}}
\def\anti{\overline}
\def\sigsi{\sigma^{\rm SI}}

\def\mw{m_W}
\def\gev{~{\rm GeV}}

\def\beq{\begin{equation}}
\def\eeq{\end{\equation}}
\def\bea{\begin{eqnarray}}
\def\eea{\end{eqnarray}}
\def\gam{\gamma}
\def\kap{\kappa}
\def\lam{\lambda}
\def\Lam{\Lambda}

\def\omghsq{\Omega h^2}

\def\br{{\rm BR}}

\def\brinv{{\rm BR}_{\rm inv}}

\def\hsm{h_{\rm SM}}

\def\to{\rightarrow}

\def\hpm{H^\pm}
\def\hp{H^+}
\def\hm{H^-}

\def\mhpm{m_{\hpm}}
\def\mh{m_h}
\def\mH{m_H}

\def\mS{m_S}
\def\lamh{\lam_h}
\def\lamH{\lam_H}

\def\hsm{h_{\rm SM}}

\def\tanb{\tan\beta}
\def\cotb{\cot\beta}
\def\sinb{\sin\beta}
\def\cosb{\cos\beta}
\def\sina{\sin\alpha}
\def\cosa{\cos\alpha}
\def\sb  {s_{\beta}}
\def\cb  {c_{\beta}}
\def\vev#1{\langle #1 \rangle}
\def\muggf{\mu({\rm ggF+ttH})}
\def\muvbf{\mu({\rm VBF+VH})}

\def\l{\left}

\def\dg{\dagger}

\def\lam{\lambda}

\def\kap{\kappa}
\def\sig{\sigma}
\def\gam{\gamma}

\def\ifmath#1{\relax\ifmmode #1\else $#1$\fi}
\def\ls#1{\ifmath{_{\lower1.5pt\hbox{$\scriptstyle #1$}}}}
\def\lss#1{\ifmath{^{\,\lower2.5pt\hbox{$\scriptstyle #1$}}}}

\def\half{\ifmath{{\textstyle{1 \over 2}}}}
\def\sqrthalf{\ifmath{{\textstyle{1 \over \sqrt{2}}}}}

\def\beq{\begin{equation}}
\def\eeq{\end{equation}}

\def\sig{\sigma}
\def\eg{{\it e.g.}}

\def\typei{Type~I}
\def\typeii{Type~II}
\def\cu{C_U}
\def\cd{C_D}
\def\cv{C_V}
\def\zpr{\mathbb{Z'}_2}
\def\zii{\mathbb{Z}_2}
\def\eq#1{eq.~(\ref{#1})}
\def\mA{m_A}

\def\lsim{\mathrel{\raise.3ex\hbox{$<$\kern-.75em\lower1ex\hbox{$\sim$}}}}
\def\gsim{\mathrel{\raise.3ex\hbox{$>$\kern-.75em\lower1ex\hbox{$\sim$}}}}

\def\bit{\begin{itemize}}
\def\eit{\end{itemize}}
\def\bec{\begin{center}}
\def\eec{\end{center}}
\def\bed{\begin{description}}
\def\eed{\end{description}}

\begin{document}

\title{Extending two-Higgs-doublet models by a singlet scalar field \\
- {\it the Case for Dark Matter} -}

\author{Aleksandra Drozd$^{1,2}$}
\email[]{aleksandra.drozd@fuw.edu.pl}
\author{Bohdan Grzadkowski$^1$}
\email[]{bohdan.grzadkowski@fuw.edu.pl}
\author{John F.~Gunion$^2$}
\email[]{jfgunion@ucdavis.edu}
\author{Yun~Jiang$^2$}
\email[]{yunjiang@ucdavis.edu}

\affiliation{(1) \,Faculty of Physics, University of Warsaw, 00-681 Warsaw, Poland}
\affiliation{(2) \,Department of Physics, University of California, Davis, CA 95616, USA}

\begin{abstract}
We extend the two-Higgs doublet models of \typei\ and \typeii\ by adding a real gauge-singlet scalar $S$ dark matter candidate (2HDMS models). We impose theoretical constraints deriving from perturbativity, stability, unitarity and correct electroweak symmetry breaking and require that the lightest CP-even Higgs, $h$, fit the LHC data for the $\sim 125.5\gev$ state at the $68\%$~C.L. after including existing constraints from LEP and $B$ physics and LHC limits on the heavier Higgs bosons. We find that these models are easily consistent with the LUX and SuperCDMS limits on dark-matter-Nucleon scattering and the observed $\omghsq$ for $S$ masses above about $55\gev$.  At lower $\mS$, the situation is more delicate. For points with $\mS$ in the $6-25\gev$ range corresponding to the CDMS~II and CRESST-II positive signal ranges,  the dark-matter-Nucleon cross sections predicted by the \typei\ and \typeii\ models more or less automatically fall within the $95\%-99\%$~C.L. signal region boundaries. Were it not for the LUX and SuperCDMS limits, which exclude all (almost all) such points in the case of \typei\  (\typeii), this would be a success for the 2HDMS models. In fact, in the case of \typeii\ there are a few points with $5.5 \gev  \lsim \mS\lsim 6.2 \gev$  that survive the LUX and SuperCDMS limits and fall within the CDMS~II 99\% C.L. signal region. Possibilities for dark matter to be isospin-violating in this 2HDMS context are also examined.


\end{abstract}


\maketitle


\section{Introduction}

One of the most important extensions of the Standard Model (SM) is the inclusion of additional particle(s) that comprise the dark matter (DM) of the Universe.  A particularly important possibility is a weakly-interacting-massive-particle (WIMP) with thermal relic density consistent with current observations. An important constraint on the WIMP scenario are limits on the spin-independent WIMP-nucleon cross section, $\sigsi$, the strongest of which are currently those of the LUX~\cite{Akerib:2013tjd} and SuperCDMS~\cite{Agnese:2014aze} Collaborations, where the LUX limit is strongest for DM masses above about $6\gev$ while the SuperCDMS limit is strongest for masses below this.~\footnote{We note that the XENON 100 limit \cite{Aprile:2012nq} is weaker than the LUX limit for all dark-matter masses and, thus, we do not reference it in our discussions.}
In combination, the LUX and SuperCDMS limits exclude the positive CDMS~II signal  observed for a WIMP with mass of $\sim$8.6\,GeV and cross-section of~$\sigsi \sim$\,$1.9\times10^{-41}{\rm\,cm}^2$ ~\cite{Agnese:2013cvt,Agnese:2013rvf} as well as the other positive hints (DAMA~\cite{Bernabei:2013xsa}, CoGeNT~\cite{Aalseth:2010vx,Aalseth:2014jpa}, and CRESST-II~\cite{Angloher:2011uu}) that support the findings of CDMS~II.  We note that
 isospin-violating DM (IVDM) scenarios \cite{Feng:2011vu} that could make the Xenon-based LUX limit consistent with the CDMS~II Silicon-based positive signal~\cite{Frandsen:2013cna,Feng:2013vod,DelNobile:2013gba,Cirigliano:2013zta} do not appear to be relevant given that the SuperCDMS Germanium-based limits require only minor rescaling~\cite{Chen:2014tka,Gelmini:2014psa}. 

In this paper we focus on a one-component DM model in which the WIMP is a singlet scalar particle that is present as part of an extended scalar sector 
of the electroweak theory. In particular, we consider two-Higgs-doublet models (2HDM) with an extra real scalar $S$
(we term the resulting models ``2HDMS") that is neutral under the SM gauge group.~\footnote{Here we will restrict ourself to the CP-conserving version of the 2HDM. However, this assumption is just to reduce the number of parameters. The analysis could  as well be performed  assuming either spontaneous or explicit violation of CP in the scalar sector.}  We introduce an extra $\zpr$ symmetry under which $S$ 
is the only odd field. Provided $S$ does not acquire a vacuum expectation value (VEV), it is  stable and  thereby a possible DM candidate. The 2HDMS then contains three CP-even states, $h$ and $H$ ($\mh\leq \mH$) from the 2HDM sector and $S$, a CP-odd state, $A$, and a charged Higgs pair, $\hpm$. The 2HDM context allows for increased flexibility for DM predictions as compared to adding an $S$ to the one-doublet SM in that either $h$ or $H$ can be identified with the observed SM-like CP-even state at $\sim 125.5\gev$ while the other CP-even state and the $A$ and $\hpm$ can provide additional channels for early-universe annihilation. Further, both of the CP-even states  contribute to DM scattering and annihilation.

That the 2HDM can provide a consistent description of all LHC observed signal strengths for either the  $h$ or $H$ identified as the observed $\sim 125.5\gev$ state (for the \typei\ or \typeii\ version of the model) is well-known~\cite{Chiang:2013ixa,Grinstein:2013npa,Coleppa:2013dya,Chen:2013rba,Eberhardt:2013uba,Craig:2013hca,Barger:2013ofa,Chang:2013ona,Cheung:2013rva,Celis:2013ixa,Wang:2013sha,Baglio:2014nea,Ferreira:2014naa,Kanemura:2014dea,Wang:2014lta,Dumont:2014wha,Ilisie:2014hea,Kanemura:2014bqa}. For simplicity, in this paper we consider only the case of  $\mh\sim 125.5\gev$. 
In the context of DM, the crucial new ingredient offered by 2HDMS is the presence of two independent Higgs portal couplings, $H_1^\dagger H_1 SS$ and $H_2^\dagger H_2 SS$, where $H_{1,2}$ are the two Higgs doublets of the 2HDM. As will be discussed in detail later, this is an important feature that makes it possible to decouple DM annihilation from DM scattering off nucleons. It also provides more freedom while trying to overcome
constraints from invisible decays of the $125.5\gev$ Higgs boson in a multiple scalar singlet extension of the SM~\cite{Drozd:2011aa} or in the two component DM scenario of \cite{Bhattacharya:2013hva}. 
The singlet extension of the 2HDM has been discussed earlier in \cite{He:2008qm,Grzadkowski:2009iz, Boucenna:2011hy, He:2011gc, Bai:2012nv, He:2013suk, Cai:2013zga, Wang:2014elb, Chen:2013jvg}. The primary new ingredient in the present paper is the inclusion of the full set of constraints on the 2HDM sector of the 2HDMS. These include: requiring consistency  with ``preLHC" constraints; an accurate fit to the combined ATLAS and CMS Higgs signal data when the $h$ is identified with the $\sim 125.5\gev$ state; and enforcing LHC limits  on the other Higgs bosons  ($H$, $A$ and $\hpm$) of the 2HDM using the procedures of \cite{Dumont:2014wha}.  With regard to the singlet sector, we derive and employ the constraints on the singlet parameters resulting from imposing perturbativity, stability, unitarity and correct electroweak symmetry breaking. As implicit from the $H_1^\dagger H_1 SS$ and $H_2^\dagger H_2 SS$ Higgs portal interactions, $SS$ annihilation via both the $h$ pole and the $H$ pole will be accounted for.

In the following section,  we will summarize the fits to preLHC and LHC data within the 2HDM context with $\mh\sim 125.5\gev$.  In Sec.~\ref{2HDMS_models}, we discuss the two-Higgs-doublets plus singlet model (2HDMS), including its general features and theoretical constraints as well as the properties of the singlet dark matter scalar. In Sec.~\ref{exp_con} we elaborate on the methodology of constraining the full 2HDMS parameter space using various experimental observations and limits when the 2HDM sector of the model is restricted to fit existing LHC data.  In Sec.~\ref{Full range results} we will present the results of our 2HDMS parameter space scan. 
There, we show that the combined LUX and SuperCDMS DM limits can only be satisfied for $\mS\gsim 55\gev$.  However, we do explore the extent to which IVDM scenarios arise in the 2HDMS case and how they come close to allowing the   CDMS~II signal to be consistent with the LUX limit. 
Section~\ref{sec:conclusions} contains our conclusions.  In appendices~\ref{vac_stab} and \ref{per_unit} we derive the constraints on the 2HDMS  from vacuum stability and unitarity, respectively.

\section{Fitting the 8 TeV LHC Higgs Signal in the 2HDM}
\label{2hdmfit}


The combined ATLAS and CMS data imply that the observed $\sim 125.5\gev$ state is quite consistent with  SM-like Higgs boson. Recent 2HDM efforts~\cite{Chiang:2013ixa,Grinstein:2013npa,Coleppa:2013dya,Chen:2013rba,Eberhardt:2013uba,Craig:2013hca,Barger:2013ofa,Chang:2013ona,Cheung:2013rva,Celis:2013ixa,Wang:2013sha,Baglio:2014nea,Ferreira:2014naa,Kanemura:2014dea,Wang:2014lta,Dumont:2014wha,Ilisie:2014hea,Kanemura:2014bqa} have thus focused on the extent to which deviations from the SM are still possible and the implications for possibly observing such deviations and/or the other Higgs bosons in future LHC running.
Of course, one must keep in mind that there is still an enhanced $\gam\gam$ signal in the ATLAS analysis whereas $\gam\gam$ rates are somewhat suppressed according to the CMS analysis and it is only the combined results that show no $\gam\gam$ enhancement.  Should an enhancement become statistically certain in future LHC runs,  this could certainly be accommodated in the 2HDM context~\cite{Ferreira:2011aa, Ferreira:2012my, Cheon:2012rh, Alves:2012ez, Altmannshofer:2012ar, Chang:2012ve, Ferreira:2012nv, Drozd:2012vf, Chen:2013kt, Celis:2013rcs}, as could a suppression, but the analysis performed in this paper would have to be revisited.  In this paper, we take the combined data at face value and employ the very recent 2HDM fits of \cite{Dumont:2014wha} keeping only points that are consistent with observations at the (rather stringent) 68\% C.L., assuming that it is the lighter $h$ that should be identified with the observed $\sim 125.5\gev$ state. To be specific, the predicted signal strengths  in the $\muggf$ versus $\muvbf$ planes for each of the $\gam\gam$, $VV$ (where $VV \equiv ZZ, WW$), $b\anti b$, and $\tau\tau$ final states  were required to have $\chi^2<2.3$  as determined using the C.L. contours established in \cite{Belanger:2013xza}.

The parameters of 2HDM can be taken to be the mixing angle, $\alpha$, that diagonalizes the CP-even scalar sector, \footnote{We follow the conventions of~\cite{Gunion:2002zf}.} $\tanb=v_2/v_1$ where $v_{1,2}=\vev{H_{1,2}}$ (with $v_1^2+v_2^2= (246\gev)^2$), the masses of the physical Higgs bosons, $\mh$, $\mH$, $\mhpm$, and $m_{12}^2$, where $m_{12}^2$ specifies the soft breaking of the $\zii$ symmetry needed to protect the 2HDM from tree-level flavor-changing neutral currents (FCNC). The most popular 2HDM's that are free of FCNC are the \typei\ and \typeii\ models. In \typei, quark masses and Yukawa couplings derive only from one Higgs doublet, conventionally chosen to be $H_2$.  In \typeii, up quarks couple only to $H_2$ and down quarks and leptons couple only to $H_1$.  The couplings, normalized to their SM values, of the Higgs bosons to vector bosons ($\cv$) 
and to up- and down-type fermions ($\cu$ and $\cd$) are functions of $\alpha$ and $\beta$ 
as given in Table~\ref{tab:couplings}; see {\it e.g.}~\cite{Gunion:1989we} for details. The \typei\ and \typeii\ models are distinguished only by the pattern of their fermionic couplings. 

\begin{figure}[h]
\begin{center}
\includegraphics[width=0.45\textwidth]{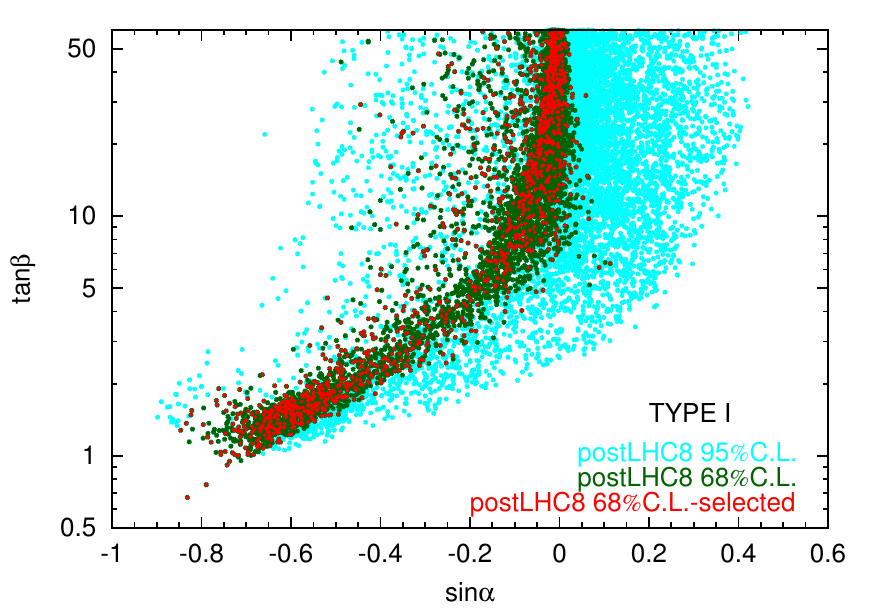}
\includegraphics[width=0.45\textwidth]{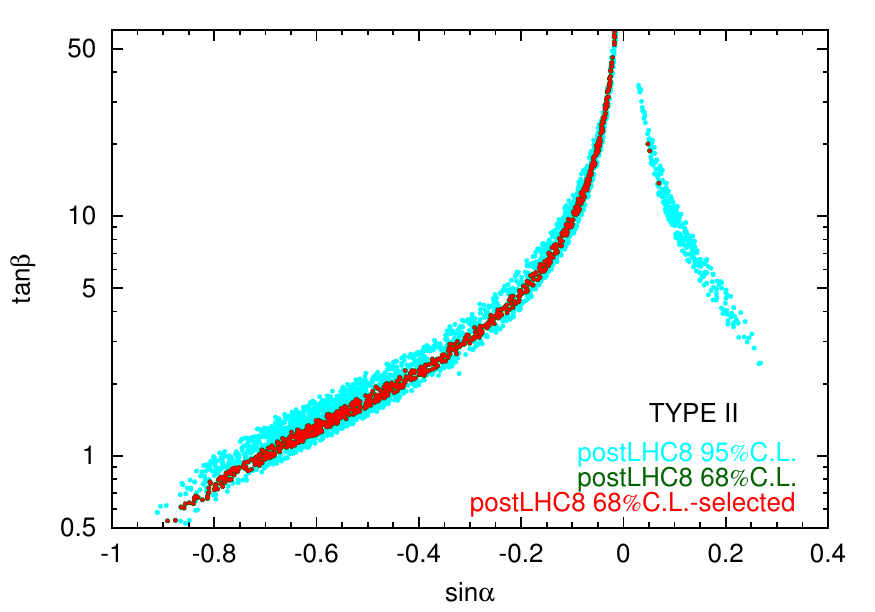}
\end{center}
\vspace{-.25in}
\caption{2HDM points in the $(\tan\beta,\sin\alpha)$ plane that provide a fit the LHC/Tevatron signal strengths at 95\%~C.L. (cyan) and 68\%~C.L. (dark green), from the analysis in~\cite{Belanger:2013xza}. In red we have marked the 68\%~C.L. points used later in the singlet scalar model analysis (for the \typeii\ model we have used all 68\%~C.L. points). }\label{2hdmsample}
\end{figure}

\begin{figure}[h!]
\begin{center}
\includegraphics[width=0.45\textwidth]{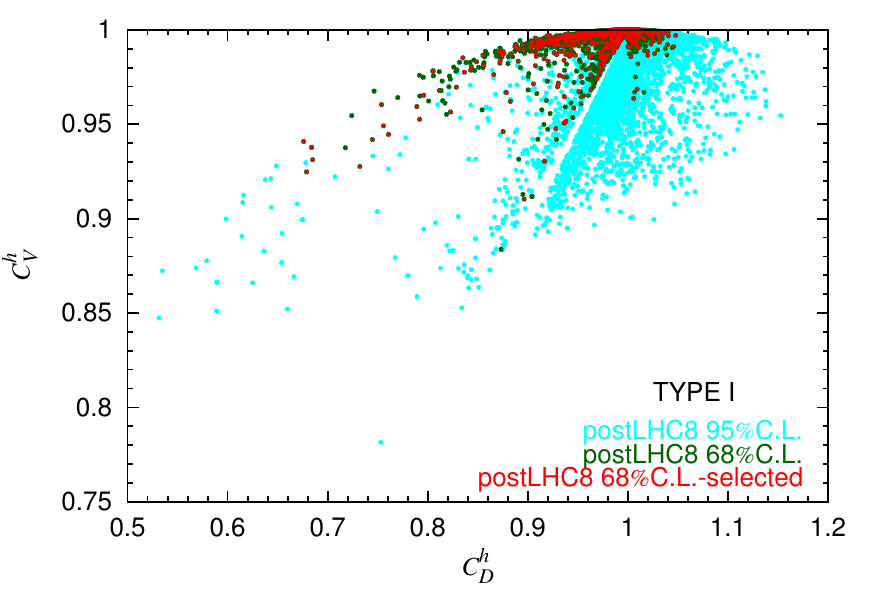}
\includegraphics[width=0.45\textwidth]{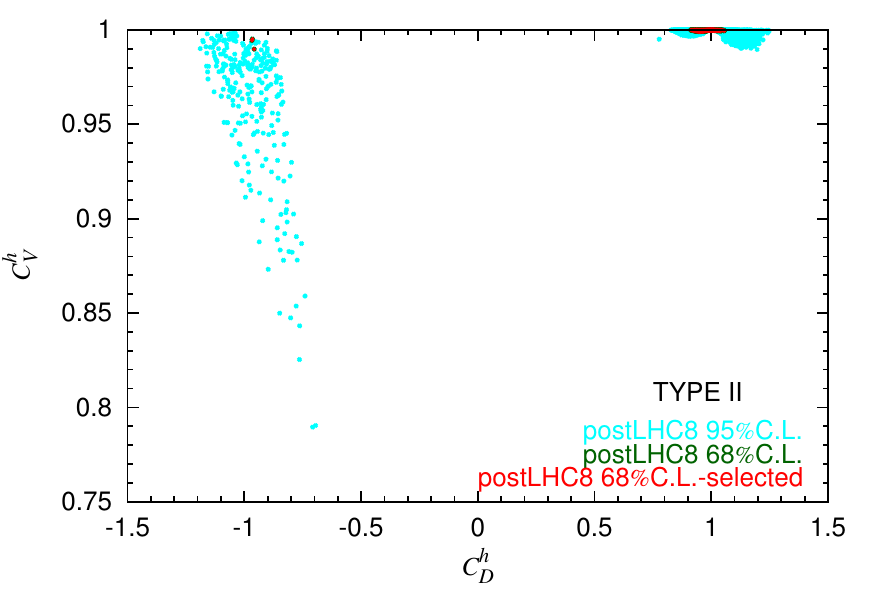}
\end{center}
\vspace{-.25in}
\caption{2HDM points in the $(C_V^h,C_D^h)$ plane using the notation of Fig.~\ref{2hdmsample}. For \typei, $C_U^h=C_D^h$.}\label{2hdmsample2}
\end{figure}

When expanding the 2HDM to include an extra singlet that could be dark matter, it is appropriate to  begin with 2HDM points that provide a good fit to the LHC data. As noted above, we assume that it is the lighter $h$ that should be identified with the $125.5\gev$ state and take the 2HDM points from \cite{Dumont:2014wha} that provide a fit to the LHC data within 68\%~C.L. These points, along with the points agreeing at the less restrictive 95\%~C.L., are shown in Fig.~\ref{2hdmsample} using the $\tanb$ vs. $\sin\alpha$ plane.  (Because there are so many 68\%~C.L. points in the \typei\ 2HDM we employ only a subset of  these points in this case --- the full 68\%~C.L. set of points are shown in dark green while the selected points are shown in red.) Of course, in order that the LHC fit for $\mh\sim 125.5\gev$ be good, the vector boson and fermionic couplings (see Table~\ref{tab:couplings}) should be quite SM-like.  The exact SM limit occurs for $\beta-\alpha=\pi/2$.
The extent to which 68\%~C.L. allows deviation in these couplings is illustrated in Fig.~\ref{2hdmsample2} where we plot the ratios of these couplings to their SM values, $C_V^h$ for the $VV$ coupling and $C_D^h$ for the down-quark. (For \typei, $C_U^h=C_D^h$.) We observe that in the case of \typeii\ almost all points have $C_V^h$ and $C_D^h$ (and $C_U^h$, not plotted) very close to unity (whereas at 95\%~C.L. significant deviations are allowed).  In the case of \typei, significant deviations in these couplings from unity are still allowed at 68\%~C.L.

\begin{table}[h!]
\begin{center}
\begin{tabular}{|c|c|c|c|c|c|}
\hline
\ & Type~I and II  & \multicolumn{2}{c|}  {Type~I} & \multicolumn{2}{c|}{Type~II} \cr
\hline
Higgs & $C_V$ & $C_U$  & $C_D$ & $C_U$ & $C_D$  \cr
\hline
 $h$ & $\sin(\beta-\alpha)$ & $\cosa/ \sinb$ & $\cosa/ \sinb$  &  $\cosa/\sinb$ & $-{\sina/\cosb}$   \cr
\hline
 $H$ & $\cos(\beta-\alpha)$ & $\sina/ \sinb$ &  $\sina/ \sinb$ &  $\sina/ \sinb$ & $\cosa/\cosb$ \cr
\hline
 $A$ & 0 & $\cotb$ & $-\cotb$ & $\cotb$  & $\tanb$ \cr
\hline 
\end{tabular}
\end{center}
\vspace{-.15in}
\caption{Tree-level vector boson couplings $C_V$ ($V=W,Z$) and fermionic couplings $C_{F}$ ($F=U,D$)
normalized to their SM values for the \typei\ and \typeii\ two-Higgs-doublet models. }
\label{tab:couplings}
\end{table}



When adding in the singlet $S$ we thus must be certain that it will not significantly disturb the fit of the $h$ to the LHC data. 
Because of the extra imposed $\zpr$ symmetry, the only influence of the $S$ on the $h$ fits arises if $h\to SS$ decays are present, which of course requires $\mS<\mh/2$. These would constitute invisible decays.  In \cite{Belanger:2013xza} a 68\% C.L. limit of $\brinv \leq 0.1$ (see also~\cite{Belanger:2013kya, Espinosa:2012vu}) was obtained in the context where the $\cu$, $\cd$ and $\cv$ coupling ratios could be varied with respect to their SM values of unity (but with $\cv \leq 1$ as appropriate to a 2HDM) and assuming no extra loop contributions to the $h\gam\gam$ and $hgg$ couplings. In the 2HDM, the $\hpm$ loops can contribute to the $h\gam\gam$ coupling, but for simplicity we will assume that $\brinv\leq 0.1$ remains applicable.  The constraint of small $\brinv=\br(h\to SS)$ plays a major role in eliminating many $\mS<\mh/2$ scenarios.

\section{2HDMS Models} 
\label{2HDMS_models}


Our goal is to analyse a model with two Higgs doublets $H_1, H_2$ and a real scalar $S$, which is a singlet under the SM gauge group. We will assign equal $U(1)_Y$ charges $Y=1$ to $H_1$ and $H_2$.  We also introduce a $\zpr$ symmetry under which $S \to - S$ (other fields are taken to be even under $\zpr$). We call this model 2HDMS.\footnote{This model was referred to as the 2HDM Darkon model (2HDMD) in some earlier literature.} The most general gauge-invariant 2HDMS scalar potential is then:
\beq
\begin{split}
V(H_1, H_2,S)&=m_1^2 H_1^{\dg} H_1+m_2^2 H_2^{\dg} H_2-\left[m_{12}^2 H_1^{\dg} H_2 + h.c.\right] +\frac{\lam_1}{2}(H_1^{\dg} H_1)^2+\frac{\lam_2}{2}(H_2^{\dg} H_2)^2+\lam_3 (H_1^{\dg} H_1)(H_2^{\dg} H_2) 
 \\
&\quad +\lam_4 |H_1^{\dg} H_2|^2 +\l[ \frac{\lam_5}{2}(H_1^{\dg} H_2)^2 + \lam_6 (H_1^{\dg} H_1) (H_1^{\dg} H_2)+\lam_7 (H_2^{\dg} H_2) (H_1^{\dg} H_2) + h.c.\right] 
 \\
&\quad
 +\frac{1}{2} m_0^2 S^2 +\frac{1}{4!}\lam_{S}S^4 +\kap_{1}S^2 (H_1^{\dg} H_1)+\kap_{2}S^2(H_2^{\dg} H_2)+S^2(\kap_{3} H_1^{\dg} H_2+h.c.)
\end{split}
\label{genpot}
\eeq
which contains 20 (real) parameters. However, for simplicity we make several additional assumptions. We consider a model without explicit CP violation (\ie\ all the $\lam$ coefficients of \eq{genpot} are taken to be real) and we only consider parameter choices for which there is no spontaneous CP breaking. As a result, the Higgs VEVs are real.  We also impose a  $\zii$ symmetry under which 
 $H_1 \to H_1, H_2 \to - H_2$, $S \to S$. This eliminates the $\lam_6$, $\lam_7$ and $\kap_3$ couplings of \eq{genpot}. However, we do allow for $m^2_{12}\ne 0$, corresponding to a soft breaking of $\zii$. The resulting potential takes the form
\beq
\begin{split}
V(H_1, H_2,S)&=m_1^2 H_1^{\dg} H_1+m_2^2 H_2^{\dg} H_2-\left[m_{12}^2 H_1^{\dg} H_2 + h.c.\right] 
 \\
&\quad +\frac{\lam_1}{2}(H_1^{\dg} H_1)^2+\frac{\lam_2}{2}(H_2^{\dg} H_2)^2+\lam_3 (H_1^{\dg} H_1)(H_2^{\dg} H_2) +\lam_4 |H_1^{\dg} H_2|^2 +\l[ \frac{\lam_5}{2}(H_1^{\dg} H_2)^2 + h.c.\right] 
 \\
&\quad
 +\frac{1}{2} m_0^2 S^2 +\frac{1}{4!}\lam_{S}S^4 +\kap_{1}S^2 (H_1^{\dg} H_1)+\kap_{2}S^2(H_2^{\dg} H_2)
\label{potsym}
\end{split}
\eeq

The next stage is to convert from the Lagrangian basis to the mass eigenstate basis.
Despite the presence of the $S^2 H_1^{\dg} H_1$ and  $S^2 H_2^{\dg} H_2$ terms, the analysis of the 2HDM sector can be performed independently of the $S$
and the usual mass matrices for the 2HDM, see \cite{Gunion:2002zf},  are not changed due to the fact that  the extra field $S$ does not acquire a VEV.~\footnote{If $S$ acquires a VEV spontaneously, as considered in \cite{Chen:2013jvg}, the $S$ mixes with the doublet Higgs and cannot be dark matter.}  However, the fields $H_1$ and $H_2$ do contribute to the $S^2$ mass term when they develop  VEVs, $H_{1,2} \to v_{1,2}$.
In terms of the mass eigenstates, the $S$-dependent part of the scalar potential has the form:
\beq
-V_S = -\frac{1}{2} m_S^2 S^2 -\lam_{h} v hS^2 - \lam_{H} v HS^2 -
S^2 (\lam_{HH}HH + \lam_{hH}hH + \lam_{hh}hh + \lam_{AA}AA + \lam_{H^+H^-}H^+H^-)
\label{V_S}
\eeq
where the physical $S$ particle mass and the DM-Higgs trilinear couplings are
\bea
m_S^2 &= & m_0^2+(\kappa_1 \cos^2\beta+\kappa_2 \sin^2\beta)v^2 \\
\lam_{h} & = & -\kappa_1 \sin\alpha\cos\beta+\kappa_2 \cos\alpha\sin\beta \label{hSS}\\
\lam_{H} & = & \kap_1\cos\alpha\cos\beta+\kap_2\sin\alpha\sin\beta \label{HSSc}\,.
\eea
While $m_0$, $\kap_1$ and $\kap_2$ constitute a complete set of extra (as compared to the 2HDM) 
free parameters for the scalar sector of the 2HDMS Lagrangian, in practice it is more convenient to employ the DM mass $m_S$ and the couplings $\lam_{h}$ and $\lam_{H}$
as the new independent set of free parameters associated with the $S$ sector.  
In the limit of $\sin(\beta-\alpha)=1$, for which the $h$ has exactly SM-like couplings to $VV$ and $f\anti f$,
\bea
\lam_{h}  &=&  \kappa_1 \cos^2\beta+\kappa_2 \sin^2\beta \label{simhSS}\,,\\
\lam_{H}  &=&  (\kappa_1-\kappa_2) \sin\beta \cos\beta\,.
 \label{simHSSc}
 \eea
We also emphasize that although there is no $ASS$ term in $V_S$ due to CP, the CP-odd Higgs boson $A$ still plays a role in determining the DM relic density through the creation/annihilation process $SS \longleftrightarrow AA$. We will discuss this issue in Sec.~\ref{exp_con}.  

The quadrilinear couplings $\lam_{HH}, \lam_{hH}, \lam_{hh}, \lam_{AA}, \lam_{H^+H^-}$ can also be expressed in terms of the $\kappa_1$, $\kappa_2$, $\alpha$ and $\beta$ parameters:
\bea
&\lam_{AA}=\half \lam_{\hp\hm}=\half(\kap_1\sin^2\beta+\kap_2\cos^2\beta)&\\
&\lam_{hh}=\half(\kap_2 \cos^2\alpha+\kap_1\sin^2\alpha)\,,\quad \lam_{HH}=\half(\kap_1\cos^2\alpha+\kap_2\sin^2\alpha)\,,\quad \lam_{hH}=\half(\kap_2-\kap_1)\sin 2\alpha\,.
\label{lamaa}
\eea
 We note that the above Lagrangian-level trilinear and quadrilinear couplings convert to Feynman rules according to:
 \beq
 g_{hSS,HSS}=-2\lam_{h,H}v,\quad g_{hhSS,HHSS}=-4\lam_{hh,HH}, \quad g_{hHSS}=-2\lam_{hH},\quad g_{AASS}=g_{\hp\hm SS}=-4\lam_{AA}\,.
 \label{gfeynman}
 \eeq
 
The fermionic couplings in the 2HDMS depend upon the behavior of the fermionic fields under $\zii$ and $\zpr$. We assume that the
fermions are even under $\zpr$ so that the $S$ has no tri-linear coupling to $f\anti f$~\footnote{We do not consider here the possibility of coupling the singlet 
to the Majorana mass term for right-handed neutrinos, $s \nu_{R\, i}^T C \nu_{R\, j}$ for $i\neq j$. In fact such couplings are
allowed if $\nu_{R\, i}$ carry $\zpr$ charge, see \cite{Grzadkowski:2009mj}.}. Fermionic couplings to $H_1$ and $H_2$ depend on the $\zii$ signs for fermions. We choose these so as to forbid flavour-changing Yukawa couplings for the neutral Higgs bosons, resulting in the couplings of Table~\ref{tab:couplings} for the models of \typei\ and \typeii. 
From now on, we restrict ourselves to the $\mathbb{Z}_2\times \zpr$ case.

 Further constraints on the model are as follows.
\subsection{Perturbativity}
\label{subsec:perturbativity}

All quartic Feynman rules associated with the mass eigenstates $h,H,A,H^{\pm},S$ are required to
satisfy the standard perturbativity constraint, \ie\ their absolute values must be  $\leq 4\pi$. 
As regards the sector involving the $S$ field, the quartic couplings of interest are those in which $S^2$ multiplies two 2HDM fields and the $S^4$ term.   One can show that the quartic Feynman rules  connecting $S^2$ to two neutral 2HDM fields, summarized above, are  guaranteed to be smaller than $4\pi$ in absolute value if  $|\kap_1|,|\kap_2|\leq 4\pi$ is imposed.   However, these maximum values are only allowed for $\alpha=\pm \pi/4$. The Feynman rule for $S^4$ interactions being $\lam_S$ means that we must also impose $0<\lam_S\leq 4\pi$, the lower bound being that required for stability.

\subsection{Vacuum Stability}

We require that the vacuum is stable at  tree level, which means that the potential in (\ref{potsym}) has to be bounded from below. As already noted this requires first of all that $\lam_S>0$. Given this, 
it is shown in Appendix~\ref{vac_stab} that  the necessary and sufficient conditions for stability read:
\bea
& \lam_1, \lam_2, \lam_S >0, \quad\lam_3+\lam_4-|\lam_5|>-\sqrt{\lam_1 \lam_2}, \quad
\lam_3>-\sqrt{\lam_1 \lam_2} & \label{first3}\\
&  \kap_1 > - \sqrt{\frac{1}{12}\lam_{S}\lam_1}, \quad 
\kap_2 > - \sqrt{\frac{1}{12}\lam_{S}\lam_2}\,. &
\label{kapsimp}
\eea
If $\kap_1~{\rm or}~ \kap_2<0$, then we have to satisfy also:
\bea
& - 2\kap_1\kap_2+\frac{1}{6}\lam_{S}\lam_3>-\sqrt{ 4\left(\frac{1}{12}\lam_{S}\lam_1 - \kap_{1}^2 \right)\left(\frac{1}{12}\lam_{S}\lam_2-\kap_{2}^2\right)}& \label{kapcomp1}\\ 
& - 2\kap_1\kap_2+\frac{1}{6}\lam_{S}(\lam_3 +\lam_4-|\lam_5| )>-\sqrt{ 4\left(\frac{1}{12}\lam_{S}\lam_1 - \kap_{1}^2 \right)\left(\frac{1}{12}\lam_{S}\lam_2-\kap_{2}^2\right)} \,.&
\label{kapcomp2}
\eea
The  conditions in \eq{first3}  above are the standard 2HDM stability conditions. These are supplemented by the  requirements of \eq{kapsimp}, \eq{kapcomp1} and \eq{kapcomp2} in the presence of the singlet field.

\subsection{$\mathcal{S}$-Matrix unitarity}
In addition, there are constraints deriving from unitarity that are closely correlated with the constraints from perturbativity.
Indeed, the dominant non-vanishing contributions to amplitudes for two-body scattering at high energy come from the processes mediated by quartic couplings. Therefore, the unitarity constraint for $J = 0$ partial waves, $|a_0|\leq 1/2$, reduces to a constraint on these quartic couplings. In Appendix~\ref{per_unit},  we describe in more detail the unitarity bounds and give explicit formulae for the scattering matrix of two-body processes in the scalar sector of the 2HDMS model.
~\footnote{It is important to note that the  $2\to 2$ scattering matrix that is obtained when  $S$-related channels are included always has a maximum eigenvalue that is larger than that of the pure 2HDM $2\to 2$ scattering matrix.  This is called the ``bordering theorem" (see \eg\ \cite{bordering}). Thus, although our 2HDM points have already been filtered using the {\sf 2HDMC} code~\cite{Eriksson:2009ws,Eriksson:2010zzb} which imposes unitarity in the 2HDM context, the unitarity limits obtained after including the $S$-related channels are guaranteed to be stronger. }

\begin{figure}[t!]
\begin{center}
\hspace*{-10mm}\includegraphics[width=0.6\textwidth]{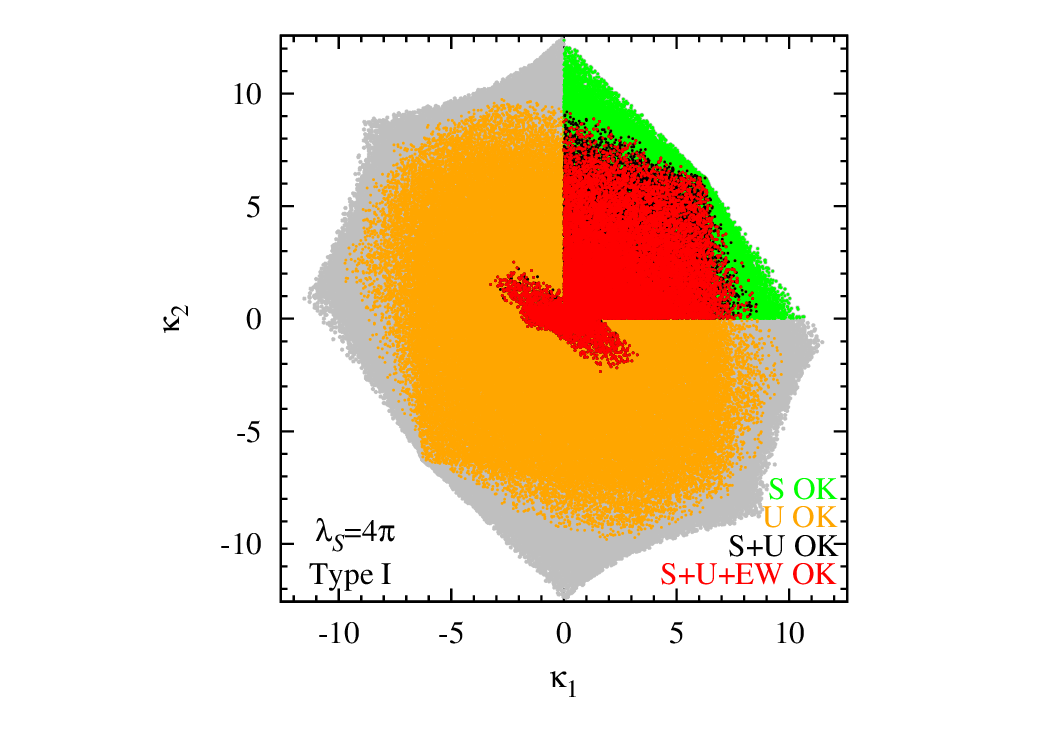}
\hspace*{-30mm}
\includegraphics[width=0.6\textwidth]{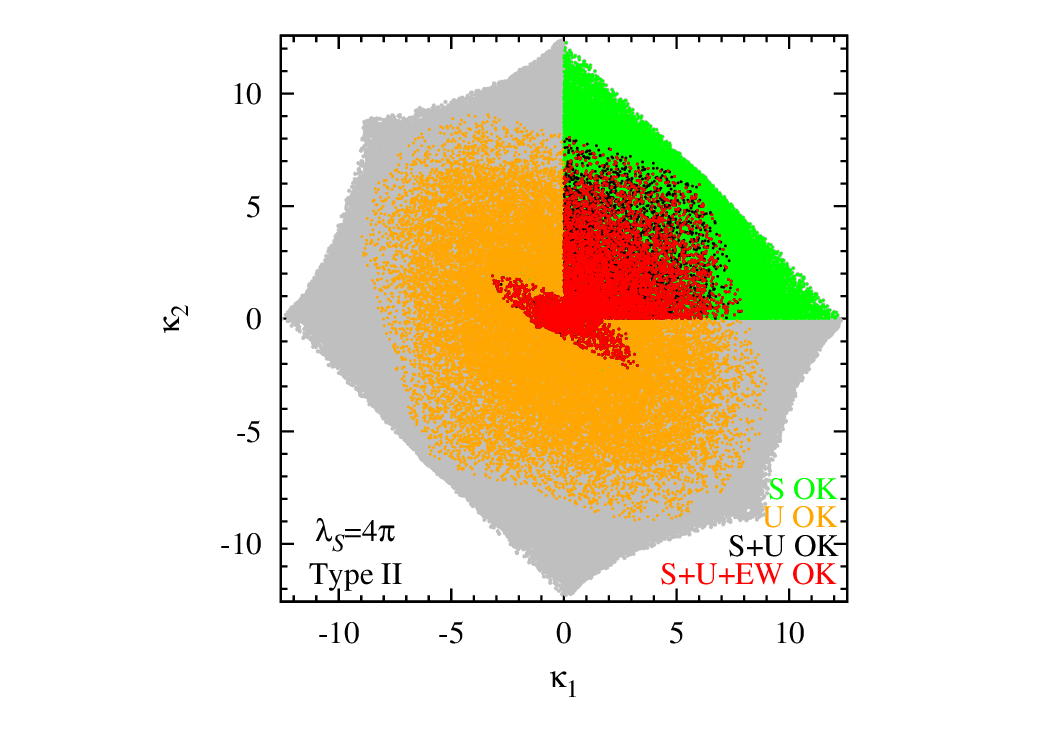}\\
\hspace*{-10mm}
\includegraphics[width=0.6\textwidth]{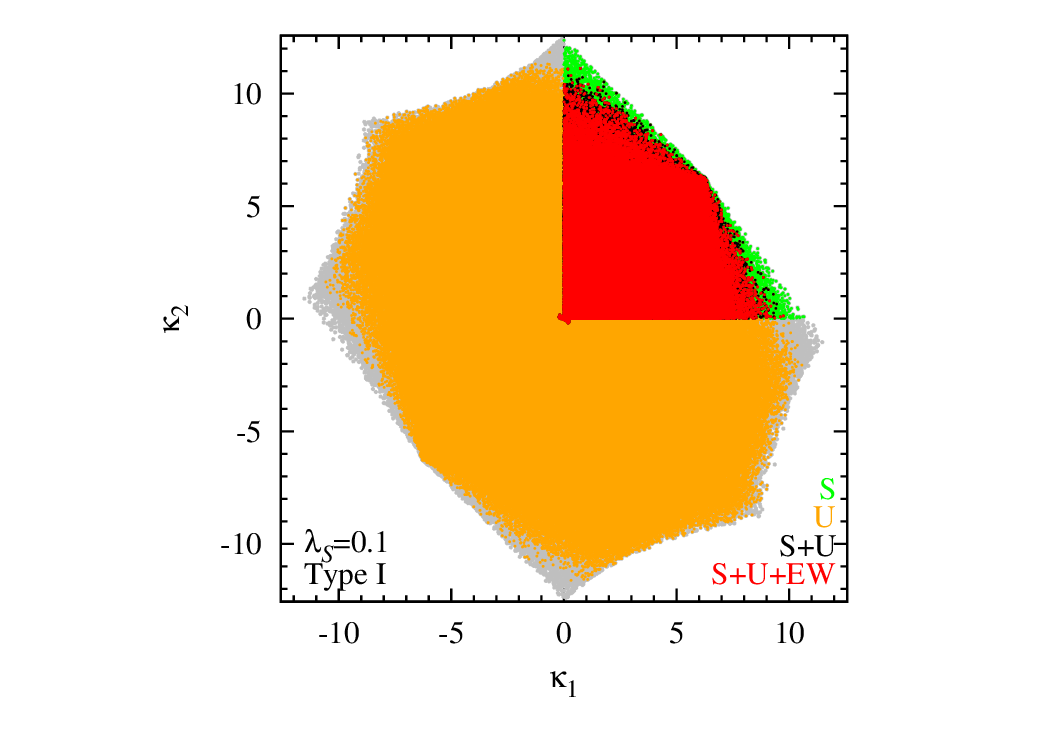}
\hspace*{-30mm}
\includegraphics[width=0.6\textwidth]{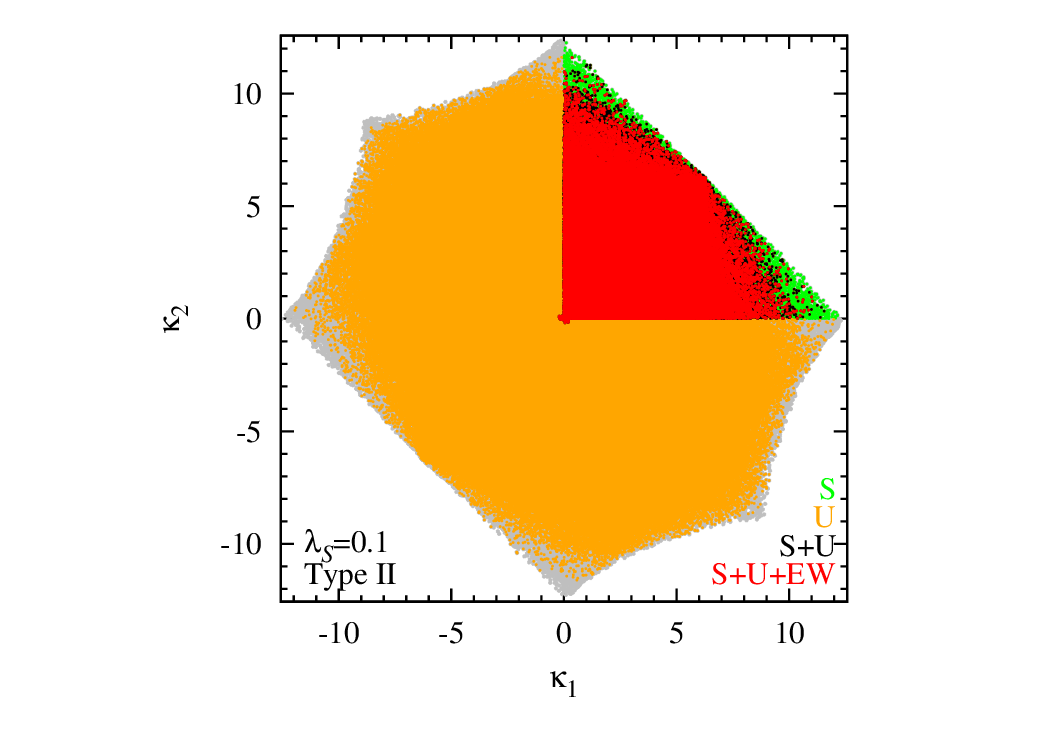}
\end{center}
\vspace{-.25in}
\caption{The plot shows the impact of the perturbativity (P), vacuum stability (S), unitarity (U)  and electroweak symmetry breaking (EW) global minimum bounds discussed in Sec.~\ref{2HDMS_models} on the $(\kap_1,\kap_2)$ plane. At the first level, the grey points are those which satisfy P --- all subsequent point layers obey P. Note that 
$|\kap_1|,|\kap_2| \leq 4\pi$ contains the perturbative region (see section~\ref{subsec:perturbativity}).  Subsequent point layers were plotted in the following order:  points after the stability bound, S (green), points after the unitarity bound, U (orange), points after the stability and unitarity bounds, S+U (black), points after the stability, unitarity and EW bounds, S+U+EW (red).  The value of the  $\lambda_S$ parameter was set to $4\pi$ ($0.1$) in the upper (lower) plots. In this figure, no restriction on $\mS$ is imposed.} 
\label{set4pi}
\end{figure}

\begin{figure}[h!]
\begin{center}
\hspace*{-5mm}
\includegraphics[width=0.55\textwidth]{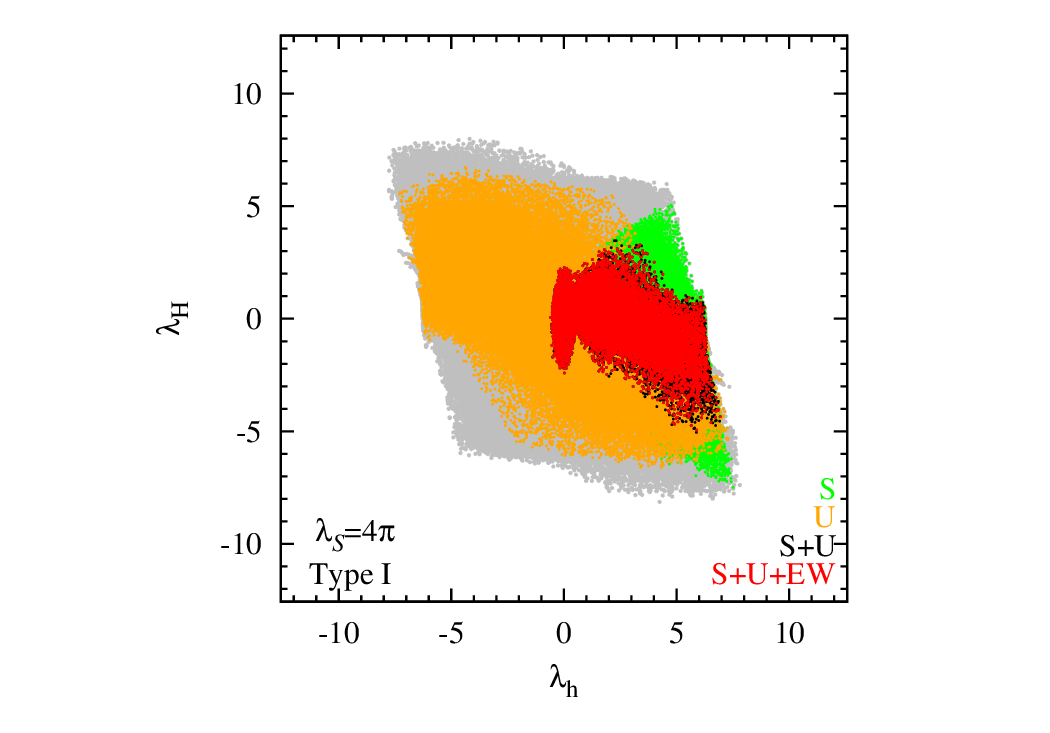}
\hspace*{-30mm}
\includegraphics[width=0.55\textwidth]{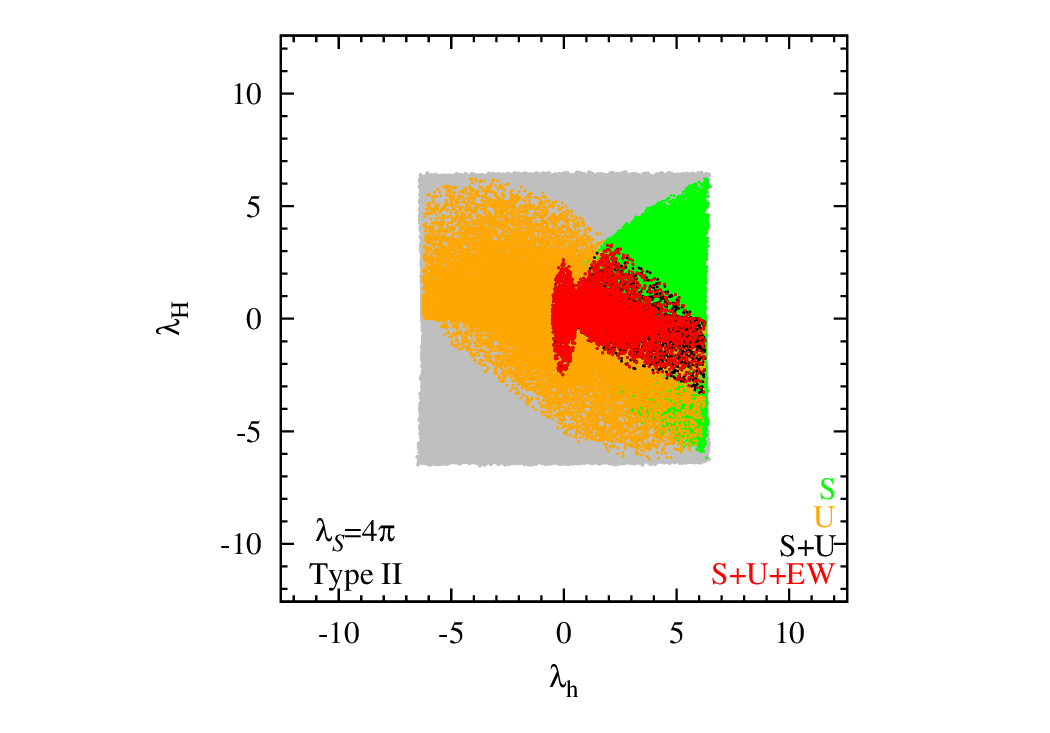}\\
\hspace*{-5mm}
\includegraphics[width=0.55\textwidth]{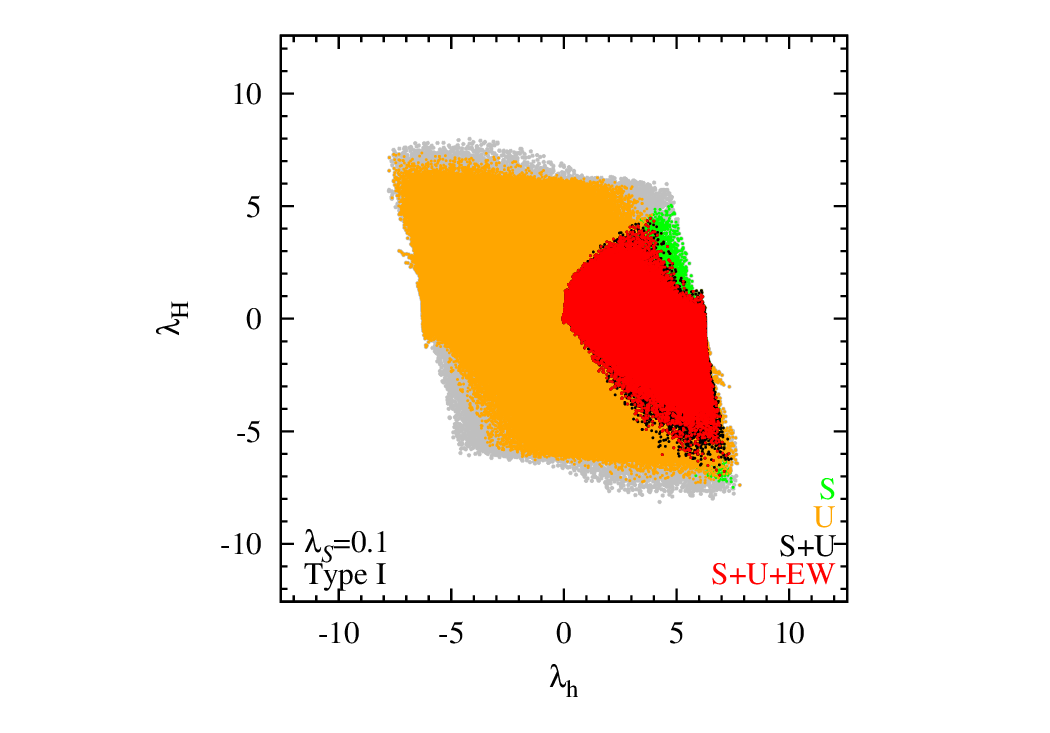}
\hspace*{-30mm}
\includegraphics[width=0.55\textwidth]{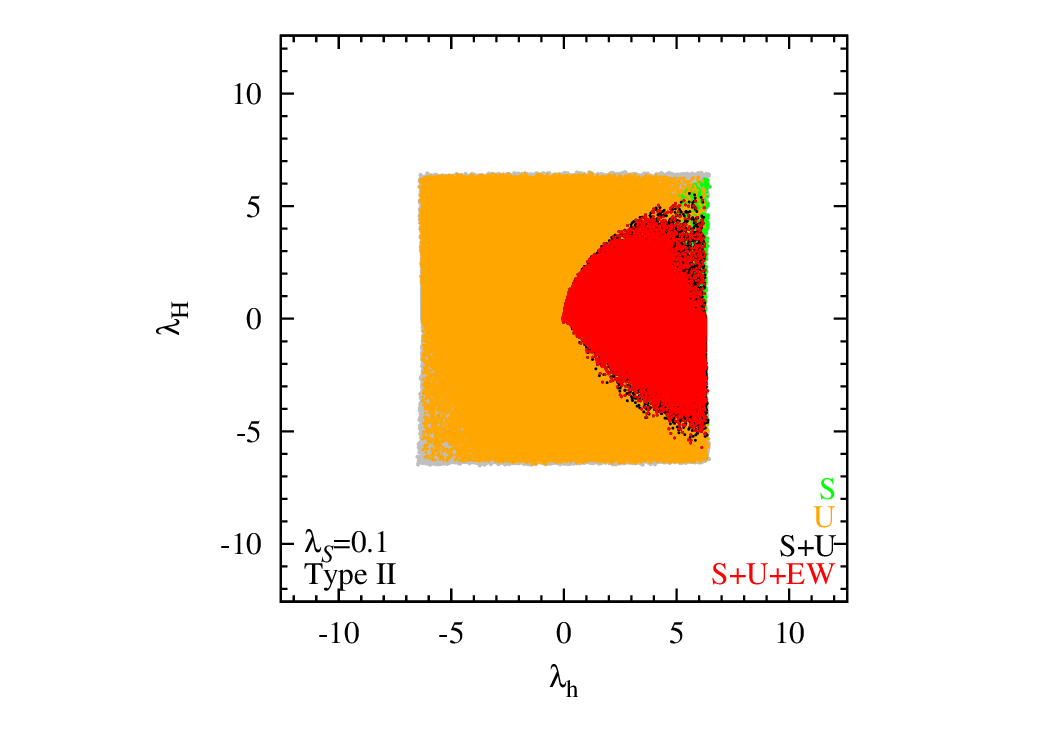}
\end{center}
\vspace{-.25in}
\caption{Bounds in the $(\lam_h,\lam_H)$ plane associated with the sequential constraints as described in the caption for Fig.~\ref{set4pi}.  No restriction on $\mS$ is imposed.}
\label{setpt1}
\end{figure}

\subsection{Electroweak Symmetry Breaking (EWSB)}

In order to ensure a stable DM particle $S$, one has to require $\langle S \rangle = 0$ at the global minimum of the scalar potential, \eq{potsym}. 
For each 2HDM point at 68\% C.L. (marked in red in Fig.~\ref{2hdmsample}), $\tanb$ and $m_{12}^2$ are given and all five $\lam$'s can be computed from the masses of the Higgs bosons and $\sina$ (see details in Appendix D of \cite{Gunion:2002zf}). With these specified, the remaining parameters $m_1$ and $m_2$ in the potential, \eq{potsym}, are determined by the minimization conditions
\beq
\begin{split}
&\quad m_{1}^2 = m_{12}^2 \tan\beta - \frac{1}{2}v^2 \left( \lambda_1 \cos^2\beta + (\lambda_3+\lambda_4+\lambda_5)\sin^2\beta \right)\\
& \quad m_{2}^2 =  m_{12}^2 \cot\beta - \frac{1}{2}v^2 \left( \lambda_2 \sin^2\beta + (\lambda_3+\lambda_4+\lambda_5)\cos^2\beta\right)\,.
\end{split}
\label{min_cond}
\eeq
Note that the minimization with respect to $S$ is trivial because of $\langle S \rangle = 0$. In practice, we  find all the minima of \eq{potsym} numerically and then 
eliminate the points for which the global minimum is not at $\langle S \rangle = 0, \langle H_1 \rangle \neq 0, \langle H_2 \rangle \neq 0$.
 


In Fig.~\ref{set4pi}, the allowed regions in the ($\kap_1, \kap_2$) parameter space are displayed after sequentially imposing the various constraints discussed above. 
\bed
\item{i)} At the first level, we impose perturbativity (P).  All subsequent point layers obey P.
\item{ii)} Next, we require vacuum stability (S).

S is always guaranteed as long as $\kap_1$ and $\kap_2$ are both positive. For $\kap_1<0$ and/or $\kap_2<0$, vacuum stability depends on the value of the $S$ self-interaction coupling $\lam_S$. 

Choosing the maximum value of  $\lam_S=4\pi$ (upper panel), there is an ellipse-shaped region of modest size where  $\kap_1$ and/or $\kap_2$ can be negative. This ellipse-shaped region  shrinks as  $\lam_S$ decreases --- we illustrate this for the case of  $\lam_S=0.1$ in the lower panel. 

\item{iii)}
Third, the unitarity conditions (U) on their own produce an oval-shaped region  in the $\kap_2$ vs. $\kap_1$ plane. 
\item{iv)}
If both S and U are imposed we are left with the black points (all red points are also black points). 
\item{v)}
Imposing S+U+EW eliminates some of the (black) S+U points, leaving us with the red points. 
\eed

\begin{figure}[h!]
\begin{center}
\hspace*{-5mm}
\includegraphics[width=0.55\textwidth]{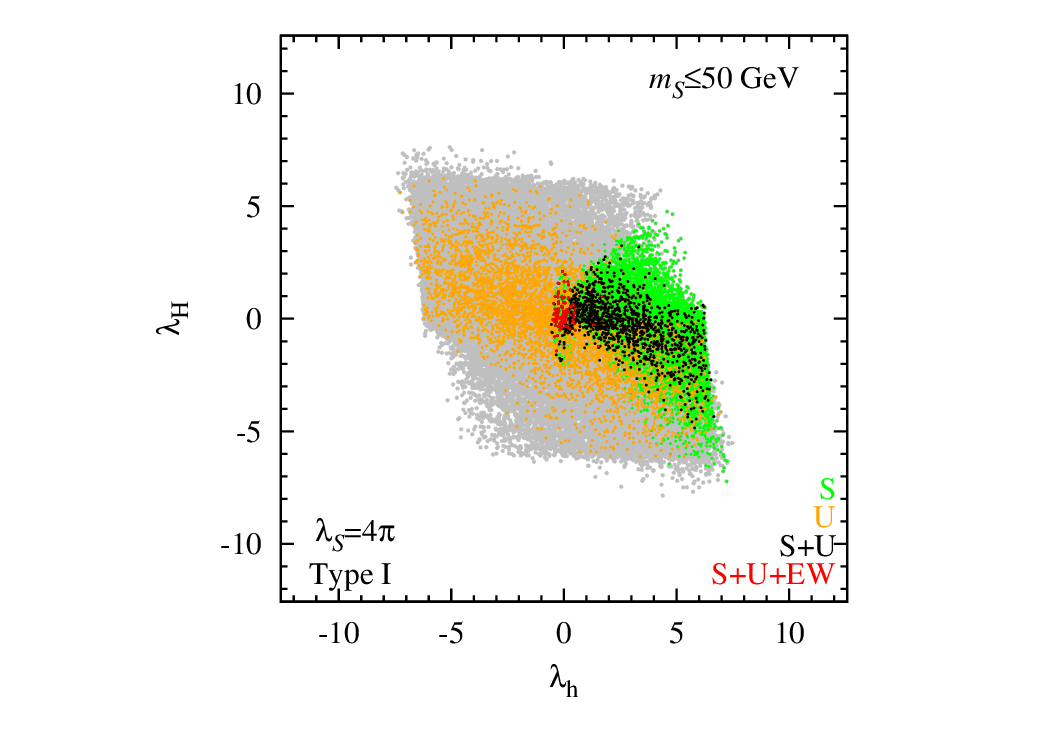}
\hspace*{-30mm}
\includegraphics[width=0.55\textwidth]{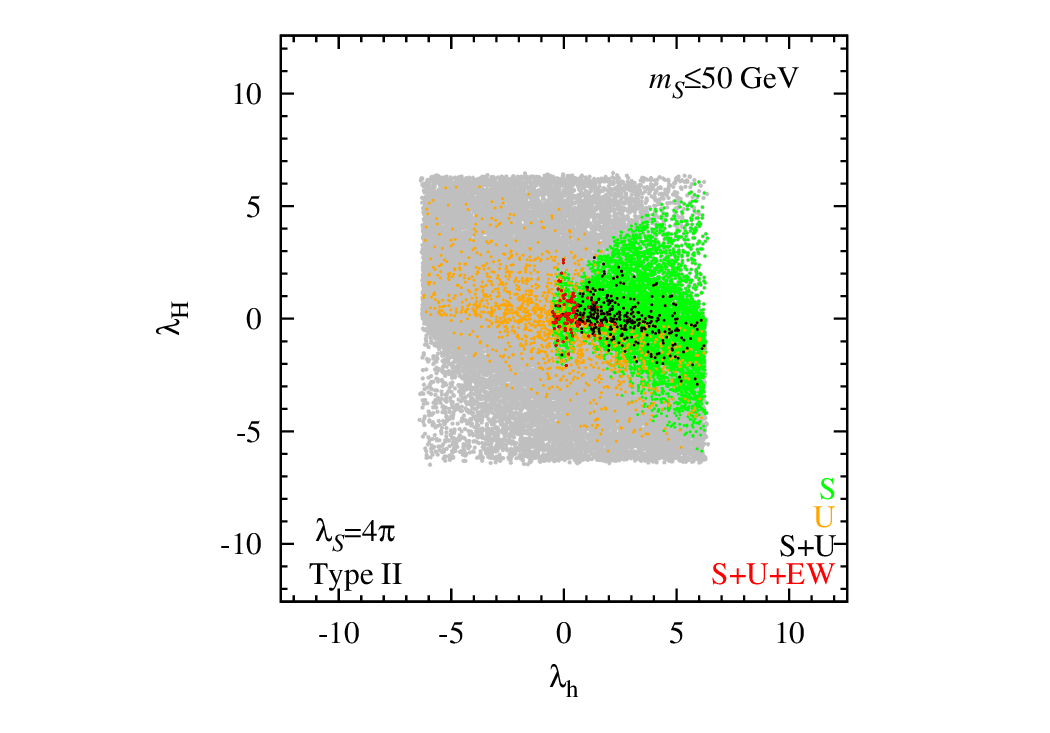}\\
\hspace*{-5mm}
\includegraphics[width=0.55\textwidth]{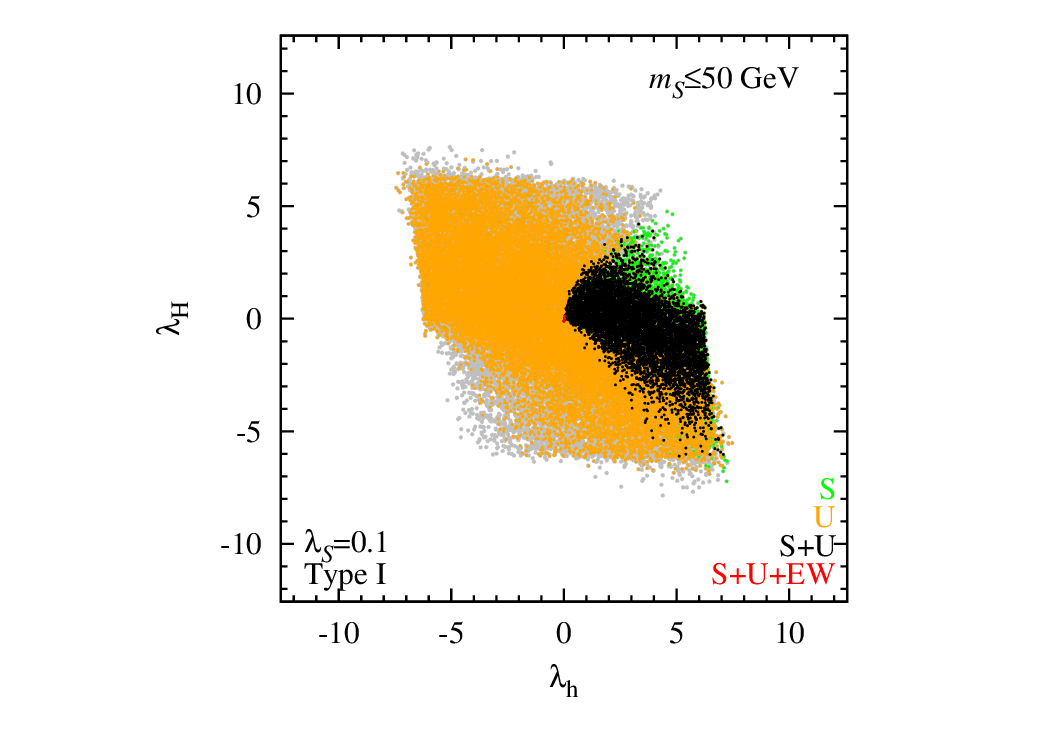}
\hspace*{-30mm}
\includegraphics[width=0.55\textwidth]{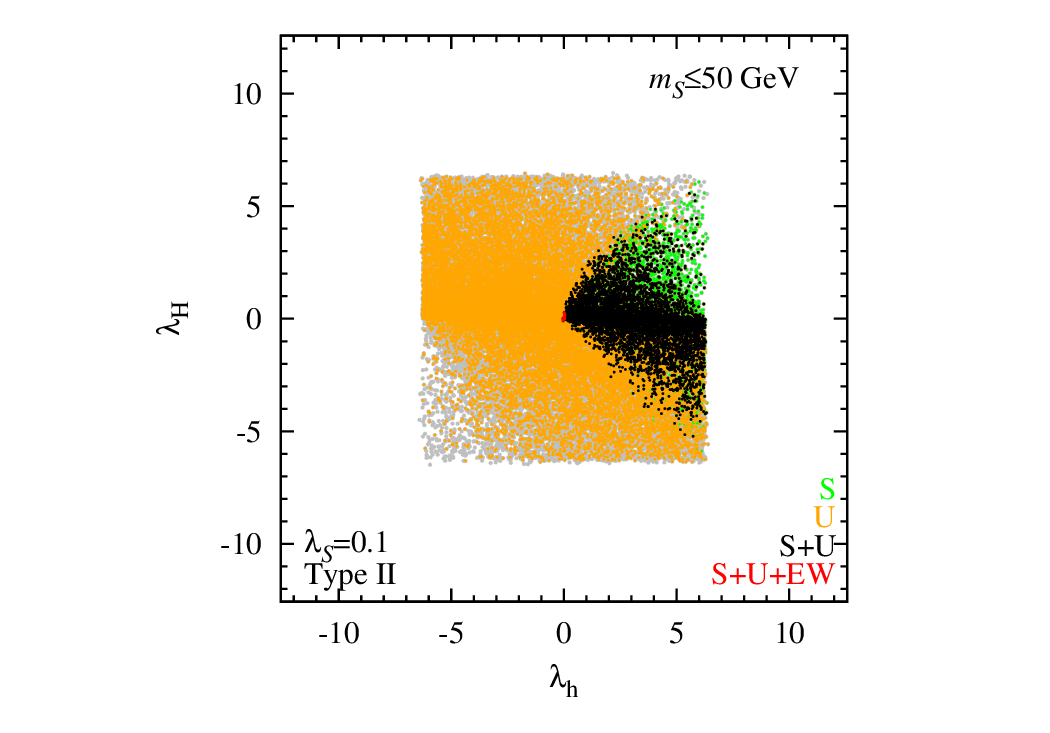}
\end{center}
\vspace{-.25in}
\caption{Bounds in the $(\lam_h,\lam_H)$ plane associated with the sequential constraints as described in the caption for Fig.~\ref{set4pi} for parameter choices yielding $\mS\leq 50\gev$. We observe that EW is an especially strong contraint in this mass region.}
\label{setsmallms}
\end{figure}

Fig.~\ref{setpt1} shows how the above $\kap_1$ vs. $\kap_2$ regions map into the $\lam_h$ vs. $ \lam_H$ parameter space.  In this figure, no restriction on $\mS$ is imposed.  In fact, the P+S+U+EW constraints are much more restrictive for $\mS<\mh/2$.  This is illustrated in Fig.~\ref{setsmallms}.  In particular, note that the maximum value of $\lamH$ that is allowed is of order $3$ in magnitude, at large $\lam_S$, and is very tiny for small $\lam_S$.  As a result, very large values of $\mH$ cannot result in sufficient annihilation through the $H$ pole diagram when $\mS<\mh/2$ given that the $h$ pole diagram is suppressed because $\lamh$ must be very small in order to avoid too large $\br(h\to SS)$.

Of course, P+S+U+EW are only the most basic constraints.  
In the following sections, we will show that once $\omghsq$ is required to agree with observations, then $|\lam_h|$ and $|\lam_H|$ are restricted to values $\lsim 0.2$ and $\lsim 2.5$, respectively.  When $m_S<\mh/2$, $\br(h\to SS)\leq 0.1$ further constrains $|\lam_h|$ to values $\lsim 0.01$. 

\vspace{5mm}


\section{Experimental constraints on 2HDMS}
\label{exp_con}

Before starting our analysis of the model, we would like to summarize the experiments that impact the extra singlet $S$ particle. 
\subsection{Dark Matter Relic Abundance}
\label{omega}

\begin{figure}[t]
\begin{center}
\includegraphics[height=2.5cm]{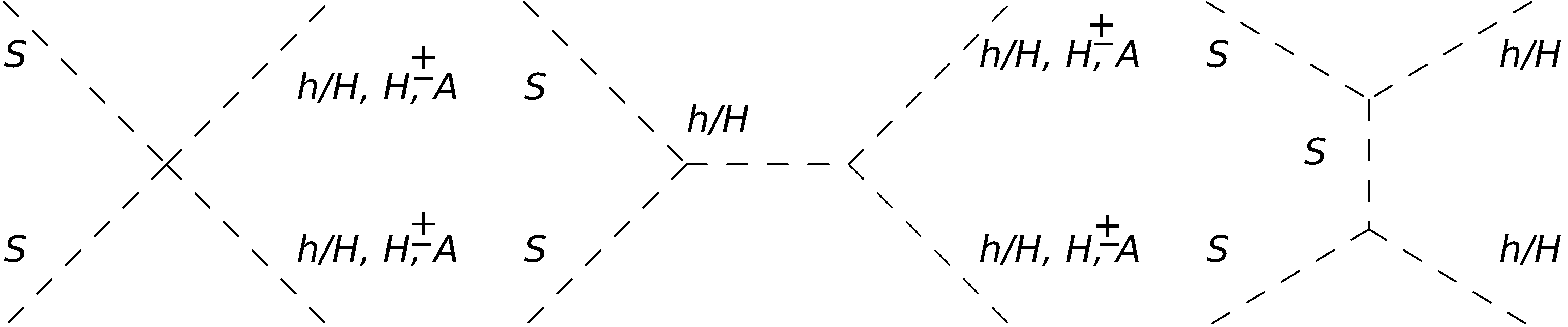}
\includegraphics[height=2.5cm]{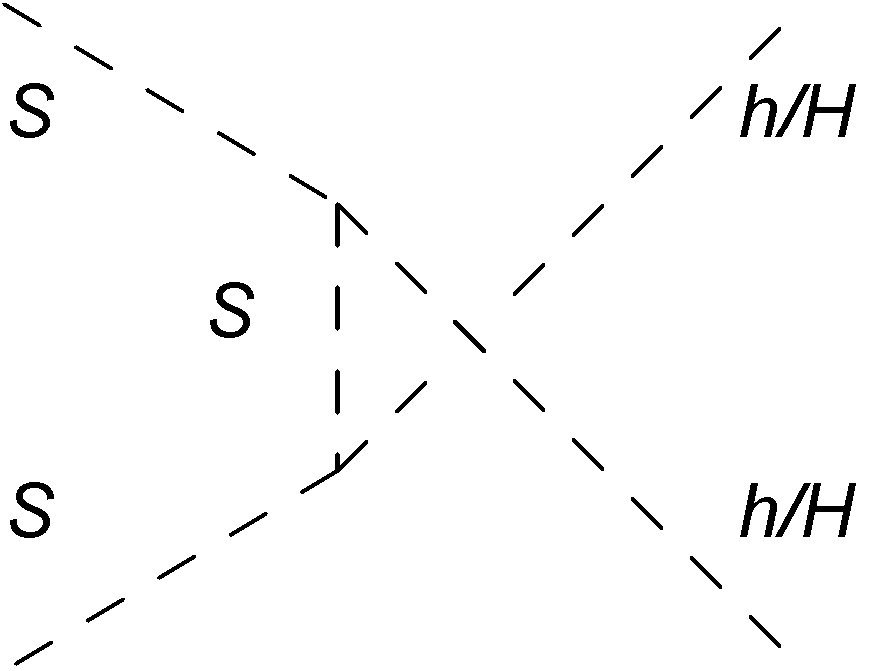}\\
\vspace{5mm}
\includegraphics[height=2.5cm]{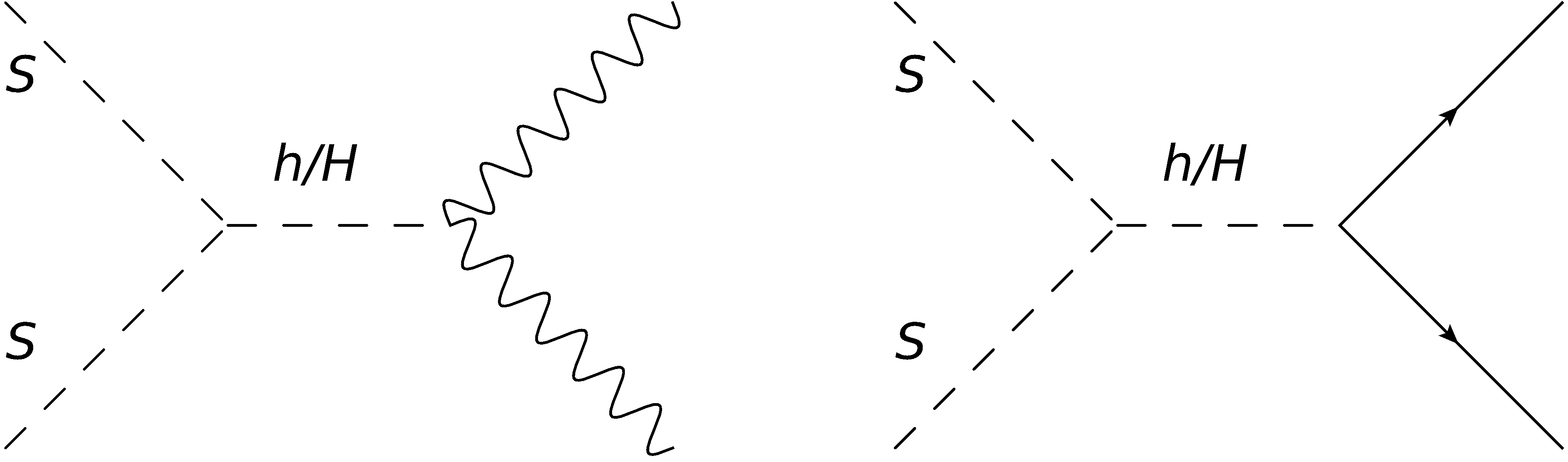}
\end{center}
\vspace{-.25in}
\caption{Singlet annihilation diagrams relevant for the relic density calculation. }
\label{RelicDiag}
\end{figure}

In the 2HDMS, the $S$ particle provides the only candidate for  DM and thus should comprise the total relic abundance of the early Universe. To a good approximation, the relic density is given by
\beq
\Omega_S 
\simeq 1.07\times10^9  \frac{x_f}{\sqrt{g_*} \, M_{\rm{Pl}}\langle \sigma_{\rm ann} v_{\rm rel} \rangle} \gev^{-1}
\eeq
where $x_f = m_S/T_f \simeq 20$ is the typical freeze-out temperature of a WIMP~\cite{Kolb:1990vq}, $M_{Pl}$ is the Planck mass, $g_*$ is number of relativistic degrees of freedom,  $\langle \sigma v \rangle$ is the thermally averaged cross section for $SS$ annihilation into the SM particles (\ie\ leptons and quarks , $f\bar{f}$, and gauge bosons, $W^+W^-, ZZ$, denoted collectively as $X\anti X$) and into Higgs bosons ($hh,hH,HH,AA,H^+H^-$). The Feynman diagrams for all the processes are shown in Fig.~\ref{RelicDiag}. First, the process of annihilation into the SM particles 
is mediated by an $s$-channel $h$ or $H$ only. Following~\cite{Grzadkowski:2009iz} (see also~\cite{Cai:2013zga}, which however has small numerical factor errors),  we find 
\beq
\langle \sigma_{SS\to X\anti X} v_{\rm rel} \rangle = \sum_{\mathcal{H}=h,H} \left| {g_{\mathcal{H}SS}C^{\mathcal{H}}_X \over 4m_S^2-m^2_{\mathcal{H}}+i \Gamma_{\mathcal{H}} m_{\mathcal{H}}} \right|^2 { \Gamma_{\text{SM}}(\mathcal{H}^* \to X\anti X) \over 2m_S} 
\label{ssxx}
\eeq
where $C_X^{\mathcal H}$ is the coupling of $\mathcal H$ to $X\anti X$ relative to the coupling of the SM Higgs boson to $X\anti X$ and 
$\Gamma_{\text{SM}}(\mathcal{H}^* \to X\anti X)$ stands for the SM partial width in the $X\anti X$ final state 
calculated at invariant mass $\sqrt{s}=2m_S$. (Note: for $X=Z$, then $\anti X=Z$ also.  In this case, $ \Gamma(\mathcal{H}^* \to X\anti X)$ must include the $1/2!$ for identical particles in the final state.)  In this equation, the  total width, $\Gamma_{\cal H}, $ must include the width for ${\cal H}\to SS$ and any partial width modifications  relative to the SM width for the various SM channels (in particular, the enhancement of $\Gamma(H\to b\anti b)$ at large $\tanb$ in the \typeii\ case.)

Second, there are all the channels containing Higgs pairs. For the $(H_i H_i)=(AA)$ or $(H^+H^-)$ final states, the relevant diagrams are the first two diagrams in the upper row of the figure, which include not only $s$-channel $h$ or $H$ exchange but also a four-point contact self-coupling. 
For final states containing CP-even Higgs pairs, $(H_i H_j)=(hh),(HH),(hH)$, there are contributions from $t$- and $u$-channel $S$ exchange (the last two diagrams with different topologies in the top row of Fig.~\ref{RelicDiag}) in addition to  the $s$-channel $h$ or $H$ exchange diagrams and the four-point contact self-coupling. 
A formula that applies to all these different cases is most easily given in terms of the Feynman rules for the various relevant vertices: the quartic Feynman rules were  given earlier in \eq{gfeynman} and the trilinear coupling $g_{{\mathcal H} hH}$ Feynman rule can be found in Appendix F of~\cite{Gunion:2002zf}. We find
\bea
\langle \sigma_{SS\to H_i H_j} v_{\rm rel} \rangle &=&{\beta(m_{H_i},m_{H_j})  \over 32 (1+\delta_{ij})\pi m^2_S} \left|  g_{H_i H_j SS} + \sum_{\mathcal{H}=h,H} {g_{\mathcal{H}SS}g_{\mathcal{H}H_iH_j}  \over 4m_S^2-m^2_{\mathcal{H}}+i \Gamma_{\mathcal{H}} m_{\mathcal{H}}} +2 \delta_{CP}{g_{H_i SS}g_{H_j SS}  \over \half(m_{H_i}^2+m_{H_j}^2)-2 m^2_S} \right|^2\,,
\eea
where 
\beq
\delta_{CP}=\left\{ \begin{array}{l l} 0 &  H_iH_j=AA,\hp\hm \\
						      1 &   H_iH_j=hh,HH,hH \\ 
						      \end{array} \right.
\eeq
and
\beq
\beta(m_{H_i},m_{H_j})=\left( 1- {m_{H_i}^2+m_{H_j}^2  \over 2 m^2_S}+{(m_{H_i}^2-m_{H_j}^2)^2\over 16 m_S^4} \right)^{1/2}\,.
\eeq
Note that some final states will typically be kinematically closed.  In particular, for $m_S <  \mh$ only the $f\anti f$ ($f\neq t$), $VV$ and, possibly, $AA$ channels will be allowed.

\begin{figure}[h!]
\begin{center}
\includegraphics[width=0.45\textwidth]{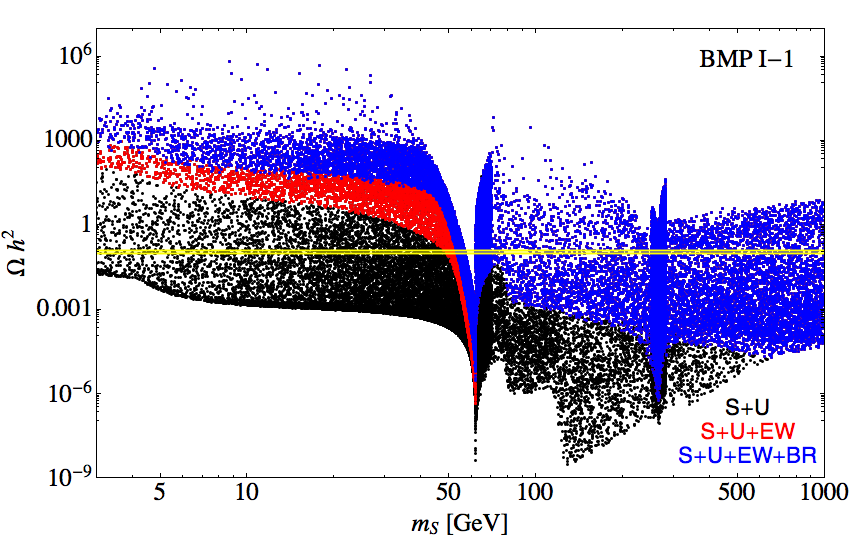}
\hspace{5mm}
\includegraphics[width=0.45\textwidth]{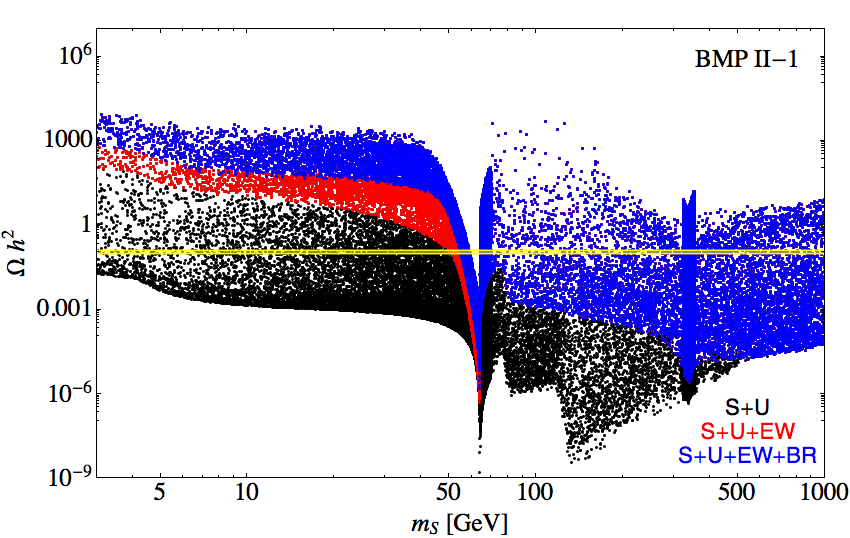}
\end{center}
\vspace{-.25in}
\caption{
Results for the relic abundance $\omghsq$ as a function of $m_S$ coming from a scan over the singlet parameter space for a fixed 2HDM point. The sample 2HDM parameters employed are given in Table~\ref{fourpoints}. All points satisfy perturbativity as defined earlier. Black points satisfy the  stability and unitarity conditions, red points satisfy also the EWSB conditions. Blue points satisfy S+U+EW {\it and} have $\br(h\to SS)\leq 0.1$. The yellow band is the recent $\pm 3\sig$ Planck window $\omghsq = 0.1187\pm 0.0017$ at 68\% CL~\cite{Ade:2013zuv}. We emphasize that the LUX and other limits on DM detection are not yet imposed in these plots.\\
}
\label{2HDMfixed}
\end{figure}

\begin{figure}[h!]
\begin{center}
\includegraphics[width=0.45\textwidth]{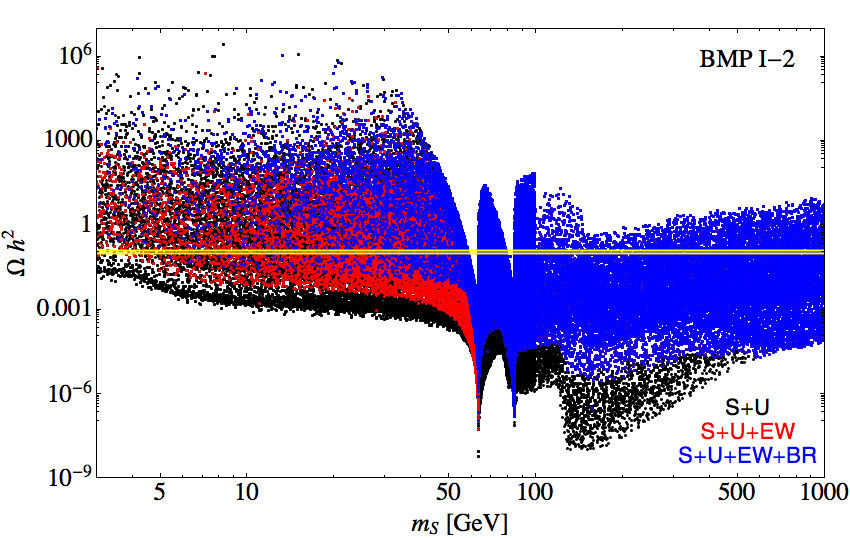}
\hspace{5mm}
\includegraphics[width=0.45\textwidth]{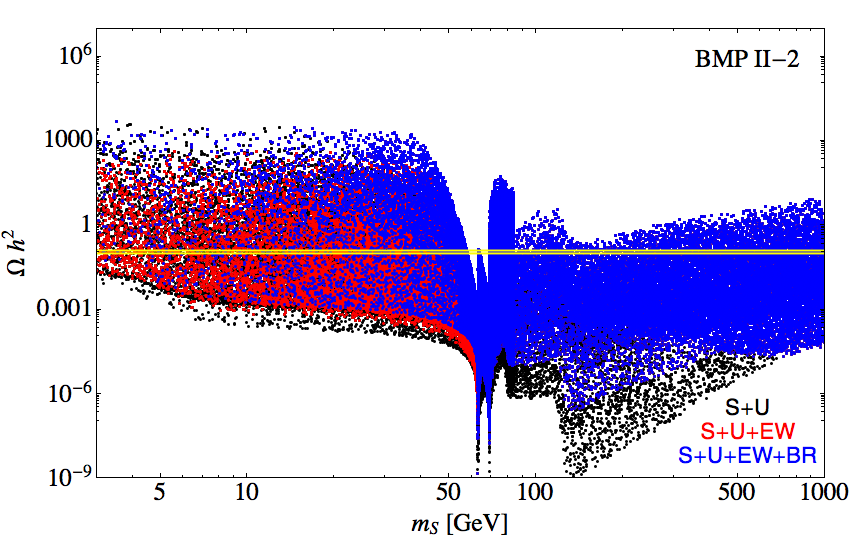}
\end{center}
\vspace{-.25in}
\caption{
As for Fig.~\ref{2HDMfixed}, but for different 2HDM points, see the last two points of Table~\ref{fourpoints},  chosen so that a large fraction of the low $m_S$ values pass all constraints other than limits on DM detection.}
\label{2HDMfixedb}
\end{figure}

In order to illustrate results of the scan over singlet parameter space, in Figs.~\ref{2HDMfixed} and \ref{2HDMfixedb} we show $\omghsq$ as a function of $m_S$ for representative 2HDM points when scanning over the remaining singlet parameters. The 2HDM parameters for these four points are given in Table~\ref{fourpoints}. For the first case, Fig.~\ref{2HDMfixed}, the 2HDM parameters are such that low $\mS$ is eliminated when correct EWSB is imposed in addition to stability and unitarity. In the second case, Fig.~\ref{2HDMfixedb}, a large range of $\mS$ values is consistent with EWSB and the observed $\omghsq\sim 0.1$. Note that for the case of Fig.~\ref{2HDMfixedb}, $\mH$ is relatively small.  This means that relatively modest values of $|\lamH|$ provide adequate annihilation for achieving the observed $\omghsq$.  In contrast, in the case of Fig.~\ref{2HDMfixed} relatively large values of $\mH$ were employed. As a result, quite large values of $|\lamH|$ would be needed for sufficient annihilation.  However, as shown in Fig.~\ref{setsmallms}, in the region of $\mS\leq 50\gev$ P+S+U+EW (especially the latter) require $|\lamH|\lsim 3$, a value that is insufficient, implying that no points satisfying P+S+U+EW (\ie\ red points) are found in this region.
In addition, at low $\mS$ values, it is possible that  $\br(h\to SS)$ is not below the 68\% C.L. upper limit of $\sim 0.1$ required by fitting of the $h$ properties to the LHC data --- see next subsection. The figures show the impact of the additional requirement of $\br(h\to SS)\leq 0.1$.


\begin{table}[h!]
\begin{tabular}{|c|c|c|c|c|c|c|c|c|c|c|c|c|c|}
\hline
BMP \# & $\tanb$ & $\sina$ & $m_{12}^2$ & $\mh$ & $\mH$ & $\mA$ & $\mhpm$  & $\lam_1$ & $\lam_2$ & $\lam_3$ & $\lam_4$ & $\lam_5$  \cr
\hline
I-1 & $1.586$ & $ -0.587$ & $5621$ & $123.71$ & $534.25$ & $645.13$ & 549.25 & 5.98 & 1.683 & 3.203 & -1.032 & -4.81 \cr
II-1 & $0.969$ & $-0.721$ & $1.251\times 10^5$ & $127.96$ & $678.98$ & $600.36$ & $563.18$ & 3.463 & 4.046 & -0.997 & -0.389 & -1.816 \cr
\hline
I-2 & $1.346$ & $-0.663$ & $-2236$ & $126.49$ & $168.01$ & $560.92$ & $556.94$ & 1.199 & 0.59 & 10.101 & -5.12 & -5.267 \cr
II-2 & $2.092$ & $-0.4096$ & $-1.264\times 10^4$ & $125.89$ & $137.86$ & $451.33$ & $398.76$ & 3.984 & 0.454 & 5.732 & -2.422 & -3.896 \cr
\hline
\end{tabular}
\caption{2HDM parameters for the plots of  Figs.~\ref{2HDMfixed} and \ref{2HDMfixedb}. Masses in GeV; $m_{12}^2$ in GeV$^2$.}
\label{fourpoints}
\end{table}

In both Fig.~\ref{2HDMfixed} and  Fig.~\ref{2HDMfixedb}, one can see a sharp dip in $\omghsq$ at $\mS\simeq 63\gev$ which arises from on-shell $h$ exchange, as well as a sudden drop in $\omghsq$ near 80/90 GeV due to the $WW$ and $ZZ$ final states becoming available in the $SS$ annihilation (the relic abundance is inversely proportional to the annihilation cross section). A  similar threshold appears around $\mS \sim m_t$. One can also observe sharp dips in $\omghsq$, corresponding to $s$-channel exchange of the heavy scalar $H$, at $m_S\simeq \mH/2\sim 265\gev$ and $340\gev$ for \typei\ and \typeii, respectively, for Fig.~\ref{2HDMfixed} and at $\mS\sim 85\gev$ and $68\gev$ in the case of Fig.~\ref{2HDMfixedb}.

\subsection{Higgs invisible/unseen decays}
\label{hig inv dec}

\begin{figure}[t]
\begin{center}
\hspace*{-.3in}
\includegraphics[width=0.45\textwidth]{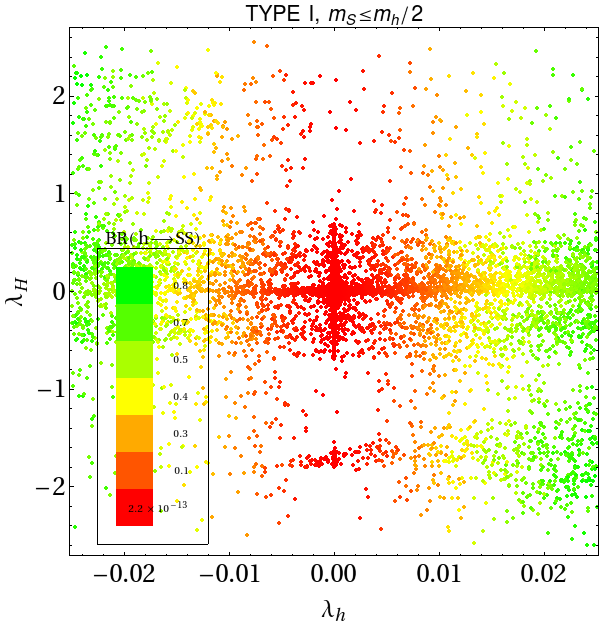}
\hspace{8mm}
\includegraphics[width=0.45\textwidth]{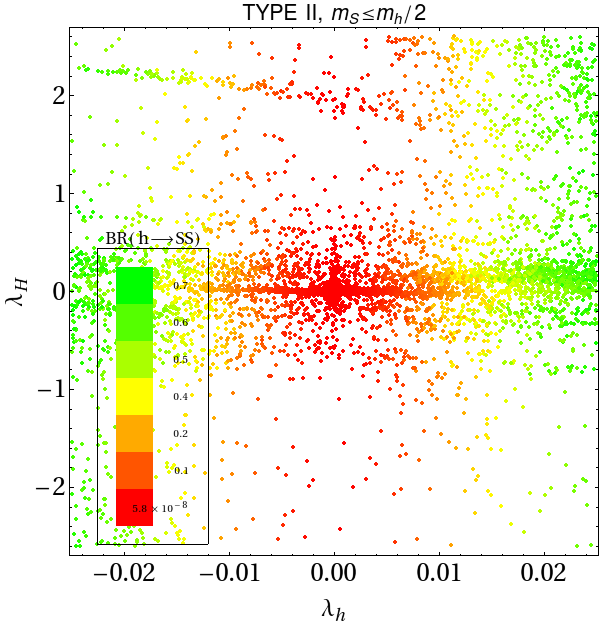}\\
\end{center}
\vspace{-.25in}
\caption{The couplings of $h/H$ to $SS$ after imposing the full set of constraints including $\omghsq$ but not LUX and  without the $\br(h\to SS)\leq 0.1$ constraint. Coloring is according to $\br(h\to SS)$: points with small $\br$ are red, large $\br$ points are green. The 2HDM points employed in this scan are the red points of Fig.~\ref{2hdmsample}.  A full scan over the singlet sector parameters is performed subject to the standard P+S+U+EW constraints. 
}
\label{lamhlamH}
\end{figure}

In addition to decays into SM particles, the CP-even Higgs bosons $h$  and $H$  of the 2HDMS have a number of possible invisible and/or ``unseen" decays.  By ``unseen" we mean decay modes that contain visible particles, but that the experimental analyses have not explored and/or are not yet able to place useful limits on. The invisible decays are $h,H\to SS$ and the potentially important unseen decay modes are $h\to AA$ and $H\to AA,hh$. Since we assume that it is the $h$ that is the $\sim 125.5\gev$ state, we are not immediately concerned with $H$ decays.  However, both $h\to SS$ and $h\to AA$ decays could make it impossible to fit the LHC Higgs data at the 68\% C.L. level that we are requiring. In fact, at this level of fitting precision, the scans of \cite{Dumont:2014wha} did not find points with $\mA<\mh/2$. Thus, we need only ensure that, for each point in the full 2HDMS parameter space,  $\br(h\to SS)$ is sufficiently small as to not significantly disturb the fit of the $h$ to the LHC Higgs data. The $h,H\to SS$ decay widths are given by:
\beq
\begin{split}
\Gamma (h_i \to SS)  = \frac{1}{2\pi} \frac{g^2_{h_i SS} }{m_{h_i}} \sqrt{1-\frac{4 m_S^2}{m^2_{h_i}}} 
\end{split}
\eeq
where $i=1,2$ denotes $h, H$ and the dimensional Feynman-rule couplings $g_{h_i SS }$ are given in \eq{gfeynman}. In what follows, it will be most convenient to discuss results in the space of the dimensionless $\lam_H$ vs. $\lam_h$ parameters, where $g_{h_iSS}=-2\lam_{h_i}v$.
When the decay $h\to SS$ is kinematically open, it will dominate the decay of the $h$ unless $\lam_{h}$ is very small. Large $\br(h\to SS)$ would invalidate the fits to the LHC $125.5\gev$ signal.  The constraints on such an invisible decay are thus quite strong:  $\br(h\to SS) \leq 10\%$ at 68\% C.L.~\cite{Belanger:2013xza}. In practice, this bound is violated for most  $\mS<55\gev$ points in the full 2HDMS parameter space leaving only a small number of points with $\lambda_{h} \ll 1$ for which $\br(h \to SS) \leq 0.1$.  This is illustrated in Fig.~\ref{lamhlamH}, which shows points in the $(\lambda_h,\lambda_H)$ plane, coloured with respect to the resulting $BR(h\to SS)$. Invisible decays of the $H$ will be discussed later.


\subsection{Direct Detection}
\label{dir det}

\begin{figure}[t]
\begin{center}
\includegraphics[width=0.2\textwidth]{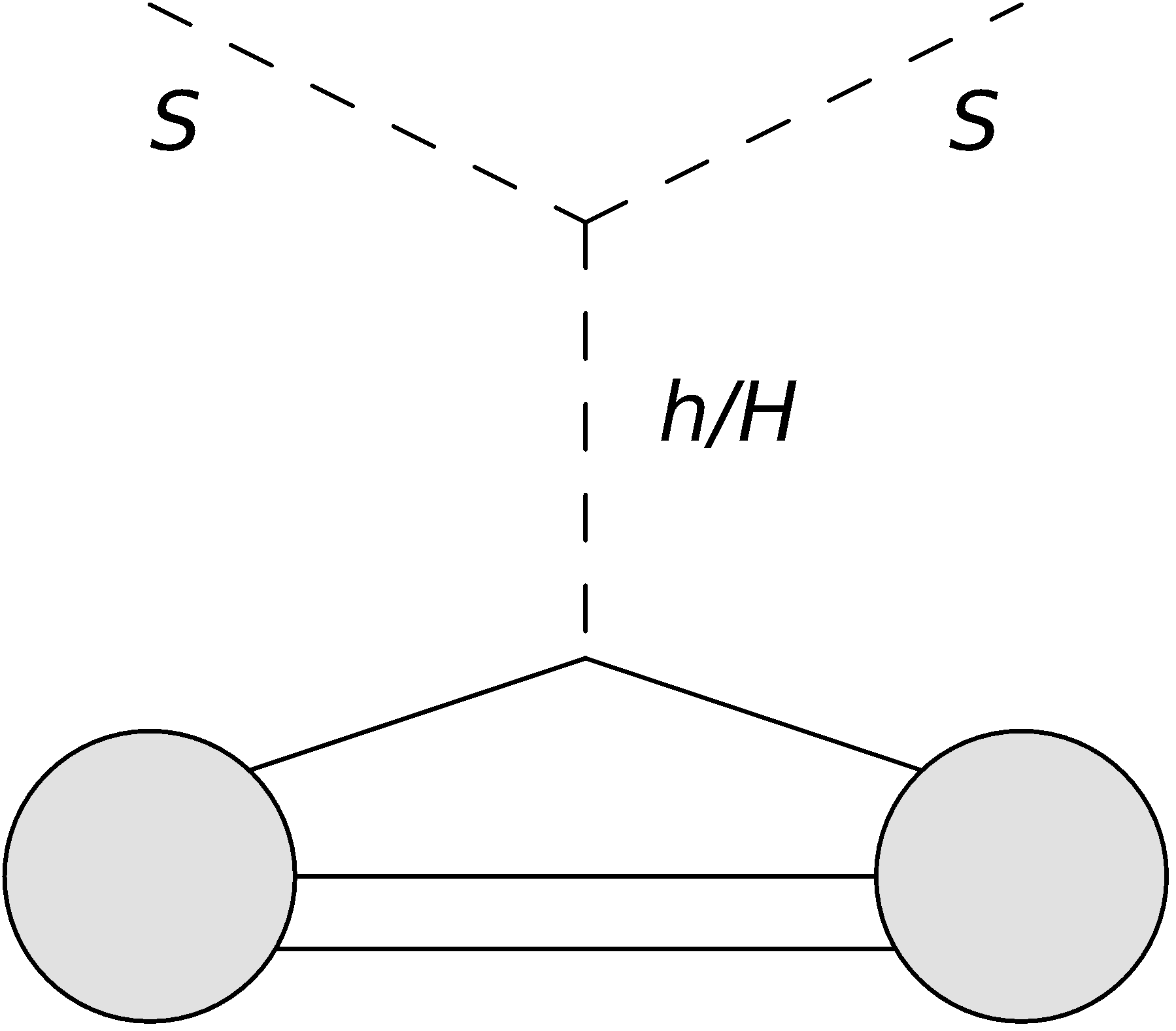}
\end{center}
\caption{Feynman diagram for the scattering of DM off a nucleon.}
\label{DDDiag}
\end{figure}

The rate at which DM-particles scattering off nuclei can be detected is directly related to the DM-nuclei
scattering cross-section~\cite{Jungman:1995df}, which is given by:
\beq
\sigma_{{\rm DM}-N} = \int_{0}^{4 \mu_r^2 v^2} \frac{d \sigma (q=0)}{d|\textbf{q}|^2} d|\textbf{q}|^2
=  \frac{4 \mu_r^2}{\pi} f_p^2 \left[Z + \frac{f_n}{f_p} (A-Z)\right]^2
\eeq
where $\textbf{q}$ is the momentum transfer, $\mu_r=(m_Nm_S)/(m_N+m_S)$ and $v$ is the relative velocity. 
The couplings of DM to the proton and 
neutron, $f_p$ and $f_n $, can be expressed as 
\beq
f_N={m_N \over 2 m_S}\left( \sum_{q=u,d,s} f^N_{Tq} {\lam_{SSqq} \over m_q} + {2\over 27} f^N_{TG} \sum_{q=c,b,t}  {\lam_{SSqq} \over m_q} \right), 
\quad f^N_{TG} = 1-\sum_{q=u,d,s}f^N_{Tq},
\quad (N=p,n).
\label{fN}
\eeq
where $m_N$ is the mass of the nucleon, $f^N_{Tq}$ is the form factor of the nucleon (see Table~\ref{formfactors}) and $\lam_{SSqq}$ is the effective coupling of the DM particle $S$ to a $q$-flavor quark component in the nucleon. In the 2HDMS, this interaction derives from  $t$-channel exchange of the $h$ and $H$, as illustrated in Fig.~\ref{DDDiag}. Thus, in the limit of zero momentum transfer, the Higgs $h_i=h$ or $H$ propagator reduces to ${i\over - m_{h_i}^2}$ and we find
\beq
\lam_{SSqq}=\sum_{h_i=h,H} {g_{h_i SS} g_{h_i q\anti q}\over -m_{h_i}^2} = \left( {2\lam_h \over m_h^2} C^h_q+{2\lam_H \over m_H^2} C^H_q \right) m_q\,,
\label{lamssqq}
\eeq
where we have used $g_{h_i q\anti q}=-i{gm_q\over 2\mw}C^{h_i}_q$ ($\mw=\half g v$ in our convention) with
the quark coupling factors $C^{h_i}_q$ for Type I and II models as listed in Table~\ref{tab:couplings} and the  Feynman rule $g_{h_i SS}$ expressions given in \eq{gfeynman}.  In practice, direct detection rates in our calculation have been evaluated using {\sf micrOMEGAs} \cite{Belanger:2013oya}, including QCD NLO corrections.

\begin{table}[h]
\begin{tabular}{c|c|c|c}
  \hline 
$ q$ &$ u$ & $d$ & $s$ \\
  \hline 
 $f^p_{Tq}$ & 0.0153 & 0.0191 & 0.0447 \cr
  \hline 
 $f^n_{Tq}$ & 0.0110 & 0.0273 & 0.0447 \cr
 \hline
\end{tabular} 
\caption{Form factors extracted from {\sf micrOMEGAs 3.0}.}
\label{formfactors}
\end{table}

There are numerous collaborations (LUX, XENON 100, SuperCDMS, CDMS, CoGeNT and DAMA being of particular interest to us) working on the direct detection of DM. They typically translate the limit on the event rate against recoil energy they directly detect into a limit on the DM-proton cross section $\sigma_{{\rm DM}-p}$ as a function of DM mass. However, in reality there are several standard assumptions hidden in this translation that might or might not be correct. For instance, they assume a DM halo in the vicinity of Earth and employ the truncated Maxwell-Boltzmann velocity distribution below the escape velocity obtained from the Standard Halo Model. They also assume that the DM particle elastically scatters with a short range contact interaction via a `heavy mediator', implyiing zero-momentum transfer. Of particular importance,  they adopt the assumption that DM has equal coupling to the neutron and proton, that is to say the ratio $f_n/f_p=1$ . 

Indeed, this equality approximately holds in the Type I model because of the universal coupling structure with up-type and down-type quarks, see Table~\ref{tab:couplings}. In fact, ${\lam_{SSqq} \over m_q}$ is independent of quark-species and the common couplings $C^h_{U,D}$ and $C^H_{U,D}$ in the \typei\ model can be factored out and will then cancel out in the ratio. From eqs.~(\ref{fN}) and (\ref{lamssqq}), one can then derive the ratio of $f_n/f_p$ in the \typei\ case: 
\beq
\frac{f_n}{f_p} =
\frac{m_n}{m_p}
\frac
{ \sum_{q=u,d,s} f^n_{Tq} + {2\over 27} f^n_{TG} \sum_{q=c,b,t}  }
{ \sum_{q=u,d,s} f^p_{Tq} + {2\over 27} f^p_{TG} \sum_{q=c,b,t}  } 
\approx 1.01208 
\label{fnfp}
\eeq
This result implies that  isospin-violating effects for DM-nucleon scattering  are negligible for a \typei\ 2HDMS  and that one can thus directly compare results of our calculations with all experimental bounds including the LUX and SuperCDMS upper limits and the CDMS~II/CRESST positive signals. 

However, the relation $f_n/f_p=1$ is not always true in the \typeii\ model. In order to compare the predicted cross-sections for DM-nucleon scattering with the results presented by the experimental groups, we define the nucleon-normalized cross section, $\overline{\sigma}_{{\rm DM}-p}$, following \cite{Feng:2011vu,Feng:2013vod}:
\beq
\overline{\sigma}_{{\rm DM}-p} = \sigma_{{\rm DM}-p} \, \Theta_X(f_n,f_p)
\eeq
where $\sigma_{{\rm DM}-p}$ is the predicted DM-proton cross-section and the rescaling factor $\Theta_X$ is defined as
\beq
\Theta_X(f_n,f_p) \equiv 
\left\{
\begin{array}{l@{\;\quad\;}l}
\left[{Z \over A}+\frac{f_n}{f_p}\left(1-{Z \over A}\right) \right]^2\, , & {\rm single\,\,isotope\,\,detector}\\[10pt]
\frac{\sum_{I}\eta_I\mu_{A_I}^2 \left[Z+f_n/f_p(A_I-Z)\right]^2} { \sum_{I}\eta_I\mu_{A_I}^2 A_I^2 }\, , & {\rm multiple\,\,isotope\,\,detector}
\end{array}
\right.
\label{DDtheta}
\eeq
where $I$ runs over all isotopes present in the detector $X$ and $\eta_I$ is the relative abundance of the $I$'th isotope. 
Note that if $f_n/f_p = 1$, then $\Theta_X(f_n,f_p) =1$. However, when $f_n/f_p \neq 1$, $\Theta_X(f_n,f_p)$ will depend upon the isotope abundances and is therefore
determined by the properties of the chemical elements used in the various detectors. It was pointed out in~\cite{Frandsen:2013cna,Feng:2013vod,DelNobile:2013gba,Cirigliano:2013zta} that the scattering amplitudes of DM with proton and neutron may interfere  destructively in such a way as to achieve   $f_n/f_p\sim-0.7$, the value for which
the resulting LUX exclusion limits  
are not in strong conflict  with the favored signal regions of the Silicon-based CDMS~II experiment and the Germanium-based CoGeNT experiment. However, these positive signal regions are in direct conflict with the limits obtained by SuperCDMS~\cite{Chen:2014tka,Gelmini:2014psa}. In any case, in order to interpret any given DM scattering result, it is necessary to compute $f_n/f_p$ for each \typeii\ parameter point.  Further, $f_n/f_p$ in general depends on the singlet sector parameters. 

However, there is an interesting special case in which $f_n/f_p$ depends only on the 2HDM parameters.
Recalling that the positive CDMS~II and CoGeNT signals are both at rather low $\mS\sim 6-12\gev$ and noting that  $BR(h\to SS)$ will be large for such masses unless  $\lambda_h$ is very small, it is useful to give an approximation for $f_n/f_p$ in the limit of $\lamh\to 0$, \ie\ in the limit of ignoring the $h$ term in \eq{fN}. In this limit, the value of $f_N$ depends only on the quark couplings of the $H$: 
\beq
f_N = {m_N\over 2\mS m^2_H} \left\{ \left[ f^N_{Tu} + {2\over 27} \left(f^N_{TGc}+f^N_{TGt} \right) \right] C^H_u+  \left( f^N_{Td}+f^N_{Ts}+{2\over 27} f^N_{TGb} \right) C^H_d \right \}, 
\quad (N=p,n).
\label{fNsimp}
\eeq

\begin{figure}[h!]
\begin{center}
\includegraphics[width=0.47\textwidth]{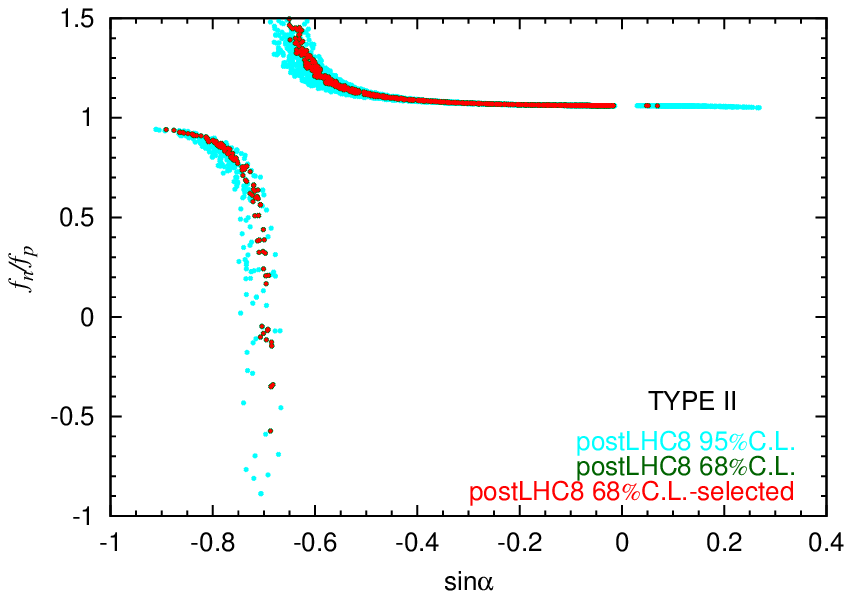}
\includegraphics[width=0.47\textwidth]{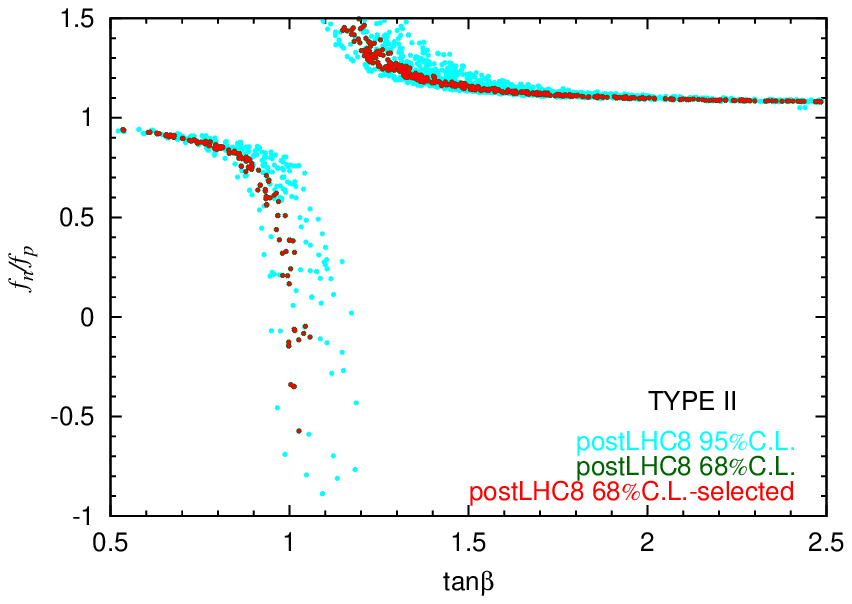}
\end{center}
\vspace{-.25in}
\caption{$f_n/f_p$ in the limit of $\lamh=0$ as a function of $\sina$ for the 68\%~C.L. \typeii\ 2HDM scan points. }
\label{onepoint}
\end{figure}

In Fig.~\ref{onepoint},  we display the resulting $f_n/f_p$ as a function of $\sina$ for the \typeii\ points from \cite{Dumont:2014wha} that give Higgs boson property fits at  the 95\%~C.L. or better.   There, we see  a large range of $f_n/f_p$ values, ranging from $+1.5$ to $\sim -0.9$.   However, for the 68\% C.L. \typeii\ points that we include in our study,  points with substantially negative $f_n/f_p$ are rather sparse, with the most negative value associated with  a single isolated point close to  $-0.7$.  This is just an accidental result given the scanning procedure/density employed in \cite{Dumont:2014wha}.  

The rather singular structure of this plot can be understood as follows. Since the LHC data at $68\%$ C.L. are  in good agreement with SM predictions, most of the \typeii\ 2HDM points shown in Fig.~{\ref{2hdmsample}} have  $\beta - \alpha \simeq \pi/2$, in which case $C^H_u \simeq -\cot\beta$, $C^H_d \simeq \tan\beta$ in the \typeii\ model. In this approximation, one can use \eq{fNsimp} to obtain  $\tan\beta$ as a function of $f_n/f_p$ in the limit of $\lam_h \to 0$:
\beq
\tan^2 \beta (f_n/f_p) ={ {f_n \over f_p}F^p_u - {m_n \over m_p}F^n_u \over {f_n \over f_p}F^p_d - {m_n \over m_p}F^n_d }
\eeq
where
\beq F^N_u \equiv f^N_{Tu} + {2\over 27} \left(f^N_{TGc}+f^N_{TGt} \right)\,, \quad F^N_d \equiv f^N_{Td}+f^N_{Ts}+{2\over 27} f^N_{TGb}\,. 
\eeq
For the value $f_n/f_p = -0.7$, one finds $\tan\beta = 1.04364$ implying  $\alpha \simeq -{\pi \over 4}$ and $\sina\sim -0.707$, with a small variation associated with the exact form factor values. Although we have  a single point with these approximate values, it turns out that for $\br(h\to SS)\leq 0.1$ the possible $\omghsq$ values lie outside the $3\sigma$ window that we have allowed.  Thus, within the limitations of the scanning so far performed we have not managed to produce a point that satisfies {\it all} the constraints not related to DM-scattering that also has $f_n/f_p\sim -0.7$, but we regard it as possible that much denser scans might reveal a point of this type.  Of course, to the extent that we accept the SuperCDMS upper bound, the CDMS II result is excluded in any case given that all the  acceptable points have $f_n/f_p$ values that are close to 1. So, it is perhaps a good feature of the 2HDMS model that obtaining a  point consistent with $f_n/f_p\sim -0.7$ and all other constraints requires a very fine-tuned choice of $\tanb$ and $\sina$.


\section{DM full mass scan}
\label{Full range results}
As noted earlier, instead of scanning over the full 2HDMS parameter space, for simplicity we used selected points from the 2HDM phenomenologically allowed points of~\cite{Dumont:2014wha} (labelled as ``postLHC8-FDOK''), as outlined in Sec.~\ref{2hdmfit}. 
In the case of $m_h\sim125.5 \gev$, the 2HDM analysis of~\cite{Dumont:2014wha} found $\sim 5200$ points consistent with Higgs observations at 68\% C.L. in the \typei\ model, from which we randomly chose 1250 points for further analysis. For the \typeii\ model we use all of the $\sim$ 900 points that fall within the 68\% C.L. criterion. These points are marked in red in Fig.~\ref{2hdmsample}. For each surviving 2HDM point, we perform a scan over the extra singlet parameters: $m_S, \lambda_h, \lambda_H$.
We then check theoretical constraints for the 2HDMS model including perturbativity, stability,  unitarity and proper electroweak symmetry breaking,  as discussed in Sec.~\ref{2HDMS_models}. Since the extra scalar $S$ does not acquire a VEV,  it does not mix with the other Higgs bosons $h$ and  $H$. As a result, the experimental constraints 
from electroweak precision tests (STU parameters), $B$ physics, direct searches at LEP and also limits on the heavier Higgs bosons ($H$ and possibly $A$)  are barely influenced by the presence of the singlet scalar $S$. Therefore, the postLHC8-FDOK points in the 2HDM can be adopted as good starting points when expanding to the 2HDMS. 
As we have noted, the only caveat that  arises is the need to take into account the possibility of $h\to SS$ decays when  the scalar $S$ is light. Substantial $\br(h\to SS)$ will spoil the pure 2HDM fit performed in~\cite{Dumont:2014wha}.  Including limits from the current Higgs invisible decay searches at the LHC   one finds roughly that $\br(h\to SS) \leq 30\%(10\%)$ is required at $95\%(68\%)$ C.L. 
Therefore, as discussed earlier, we impose a cut of $BR(h\to SS)\leq10\%$ for all points presented in the following context (except for a few situations as described later) in order to maintain the LHC signal fit and consistency with invisible decay limits.  
Finally, we use {\sf micrOMEGAs}~\cite{Belanger:2013oya} to calculate the relic abundance of the DM candidate $S$ and require that the predicted $\omghsq$ fall within the $\pm 3\sigma$ Planck window $\Omega_{\rm DM}^{\rm exp} = 0.1187\pm 0.0017$ at $68\%$ C.L.~\cite{Ade:2013zuv}. Hereafter, we refer to this set of constraints as the ``{preLUX}'' constraints.

Let us now turn to the issue of DM scattering on nuclei. For the points satisfying the ``preLUX'' constraints, we calculate the cross section for the  scattering of the $S$ off a nucleon and compare the predicted value $\overline{\sigma}_{{\rm DM}-p}$ (after rescaling by $\Theta$ in the case of \typeii) to the latest LUX limits for  the DM-proton cross section, denoted $\sigma_{{\rm DM}-p}^{\rm LUX}$ (which are obtained assuming $f_n/f_p=1$).  
If the points obey the condition $\overline{\sigma}_{{\rm DM}-p} \leq \sigma_{{\rm DM}-p}^{\rm LUX}$, they are not excluded by the LUX limit. 

\subsection{\typei\ Analysis}

\begin{figure}[h!]
\begin{center}
\includegraphics[width=0.67\textwidth]{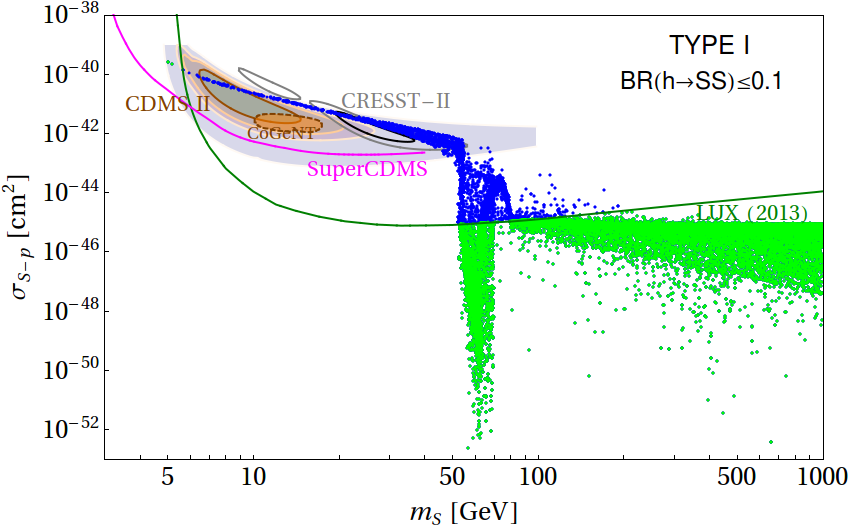}
\end{center}
\vspace{-.25in}
\caption{
Cross section for DM-proton scattering for the \typei\ model. All  points shown satisfy the full set of preLUX constraints, including $\br(h\to SS)\leq 0.1$, while the green points satisfy in addition the LUX limits. The pink
and green lines are the limits from SuperCDMS
and LUX, respectively. Recall that for \typei, $f_n/f_p\sim 1$ and so no rescaling is required between target types. Also shown are contours corresponding to the CRESST-II, CoGeNT and CDMS~II positive signal regions. In the case of CRESST-II, the darker black contour is at 68\%~C.L. and the lighter grey contours are at 95\%~C.L.  In the case of CoGeNT (orange region) we show only the 90\%~C.L. contour. For CDMS~II, we display contours (using various levels of grey) at 68\%, 90\%, 95\% and 99\% C.L.}
\label{CS_ms_all_typeI}
\end{figure}

In Fig.~\ref{CS_ms_all_typeI} we present the cross section versus  $\mS$  for the \typei\ model. Since $f_n/f_p\sim 1$ in the case of the \typei\ model, all experimental results can be displayed on the same plot. Points obeying the LUX limit are shown in green. Points that do not pass the LUX limit but do satisfy all preLUX conditions (including correct $\omghsq$ and $\br(h\to SS)\leq 0.1$)  are shown in blue. Note that few green points at very low $\mS$ that pass the LUX limit are excluded by the SuperCDMS limit.
Note that the \typei\ predictions for $\overline{\sigma}_{{\rm DM}-p} $ agree pretty well with CDMS~II/CRESST-II data (for more detailed discussion, see Sec.~\ref{low_mass}, but, of course, disobey the LUX limit.   The narrowness of the $\overline{\sigma}_{{\rm DM}-p} $ band at low $\mS$ can be understood as follows. In this mass region, we know that  $\lambda_h\simeq 0$, DM annihilation and scattering off nucleons are  thus realized via  $H$ exchange in the $s$- and $t$-channels, respectively.  Both processes are essentially controlled by the ratio  $\lambda_H/\mH^2$.   We observe that once the constraints of $BR(h\to SS) \leq 10\%$ and good $\omghsq$ are both satisfied $\lambda_h$ and $\lambda_H$ are roughly fixed. As a result, the predicted value of $\sigma_{S-p}$ as a function of $\mS$ is constrained to a narrow band that  happens to pass through the CDMS~II/CRESST-II preferred regions.  However, the CDMS~II/CRESST-II regions are simply not consistent with the combination of LUX and SuperCDMS limits in the \typei\ model.  Finally, once $\mS \gsim 55\gev$ essentially all of the points that are consistent with preLUX constraints also pass the LUX limit (SuperCDMS limits do not extend to masses $\gsim 40 \gev$).

\begin{figure}[h!]
\begin{center}
\includegraphics[width=0.4\textwidth]{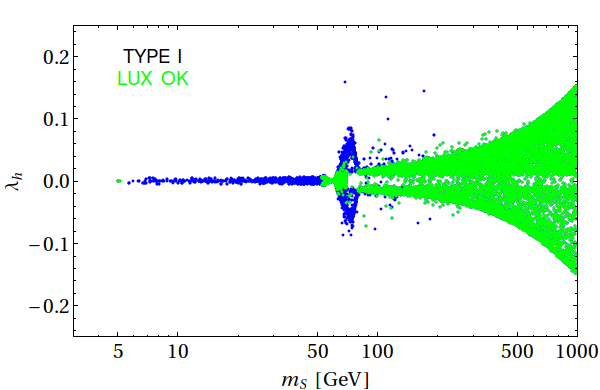}
\includegraphics[width=0.4\textwidth]{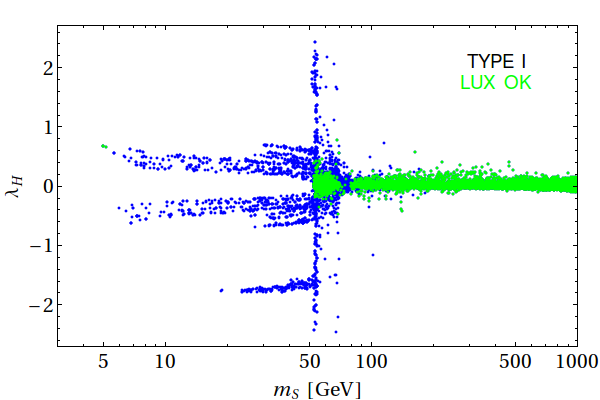}
\end{center}
\vspace{-.25in}
\caption{
The couplings $\lambda_h$ and $\lambda_H$ as a function of $m_S$ for \typei. All  points shown satisfy the full set of preLUX constraints;  blue points are excluded by LUX while green points are allowed by LUX results. The green points at very low $\mS$ are, however, excluded by SuperCDMS. 
}
\label{lamhH_ms_typei}
\end{figure}

In Fig.~\ref{lamhH_ms_typei}, we display  the associated $\lamh$ and $\lamH$ values as a function of $\mS$. 
We see that for $m_S\lsim 50\gev$, the restriction to small $\lamh$ coming from requiring $\br(h\to SS)\leq 0.1$ implies that $SS\to H\to SM$ will be dominant and correct $\omghsq$ then requires relatively substantial $\lamH$, the precise value depending on $\mH$, see \eq{ssxx}.  
In contrast, there is a considerable variety of possibilities for $\lamh$ and $\lamH$ in the ``resonance" region, \ie\ in the vicinity of $\mS\sim \mh/2$.  Typically, both the $h$ and $H$ $s$-channel diagrams contribute to $SS\to X\anti X$. Once $\mS$ is above the resonance region, many channels open up and $\lamh$ is no longer restricted by a limit on $\br(h\to SS)$.  A significant range of  $\lamh$ becomes possible, the larger the value of $\mS$ the larger the range.  Note that only in the resonance region are large values of $\lamH$ possible.  There, contributions from $h$ and $H$ exchange can partially cancel.  The fact that neither $\lamh$ nor $\lamH$ can be very large above the resonance region reflects the large number of final states that become available, in particular the $hh$ channel opens up once $\mS\gsim \mh$.

We note that the ``band" structure in the $\lamH$ vs. $\mS$ plot in the $\mS\lsim 50\gev$ region is due to the fact that $H$ exchange is dominant for $SS\to X\anti X$ annihilation.  One finds that each band is associated with a particular $\mH$ value for the associated 2HDM point. As expected from \eq{ssxx}, the larger the value of $\mH$ the larger the value of $\lamH$ that is needed for correct $\omghsq$.

\subsection{\typeii\ Analysis}

\begin{figure}[h]
\begin{center}
\includegraphics[width=0.7\textwidth]{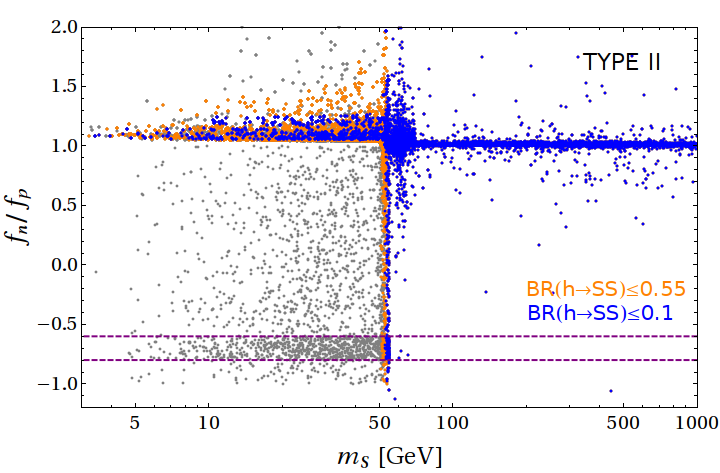}
\end{center}
\vspace{-.25in}
\caption{
We display the correlation between the ratio of $f_n/f_p$ and $\br(h\to SS)$ for points that obey all the preLUX constraints
(\ie\ theoretical constraints for 2HDMS, 2HDM fitting and correct $\Omega_{DM}$).   
Blue points have $\br (h\to SS)\leq 0.1$ (therefore blue points satisfy the full set of preLUX constraints). The purple dashed line is located  at  $f_n/f_p=-0.7 \pm 0.1$. Most grey points have $\br(h\to SS)\simeq 1$.
}
\label{BRh_fnfp_all_typeII}
\end{figure}

We now turn to the \typeii\ model. A particularly interesting question is whether or not one can have consistency between the CDMS~II/CRESST-II preferred regions and the LUX limits. As already noted, this  requires $f_n/f_p\sim -0.7$. As a first step, we  examine the correlation between the ratio of $f_n/f_p$ and $\br(h\to SS)$, as illustrated in Fig.~\ref{BRh_fnfp_all_typeII}. After imposing the constraint $\br(h\to SS) \leq10\%$, as well as all the other preLUX constraints, all points with $f_n/f_p\sim -0.7$  in the low $m_S$ region are excluded. Indeed, in the low $\mS$ region $f_n/f_p\gsim 1$. Even relaxing the invisible decay  limit to $BR(h\to SS)\leq 55\%$ (the most conservative upper bound on $\br_{\rm inv}$ at the LHC~\cite{Chatrchyan:2014tja}) still does not allow for points with  $f_n/f_p \sim -0.7$. In the resonance region of $\mS \simeq 55\gev$, a predicted  $f_n/f_p$ values range from below $-1$ to above $2$, although the majority of points have $f_n/f_p$ near 1.  Above the resonance region, \ie\ $\mS\gsim \mh/2$, most points have $f_n/f_p\sim 1$, but there is a handful of points with $f_n/f_p$ values both substantially above 1 and substantially below 1 (a few points have quite negataive values).  Thus, in our predictions for DM scattering, it will be important to take into account the variation of $f_n/f_p$.


\begin{figure}[h]
\begin{center}
\includegraphics[width=0.47\textwidth]{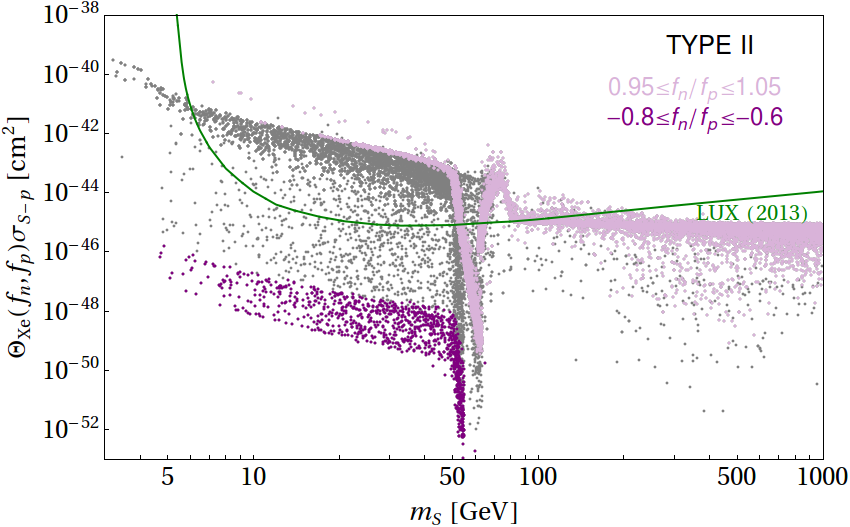}
\includegraphics[width=0.47
\textwidth]{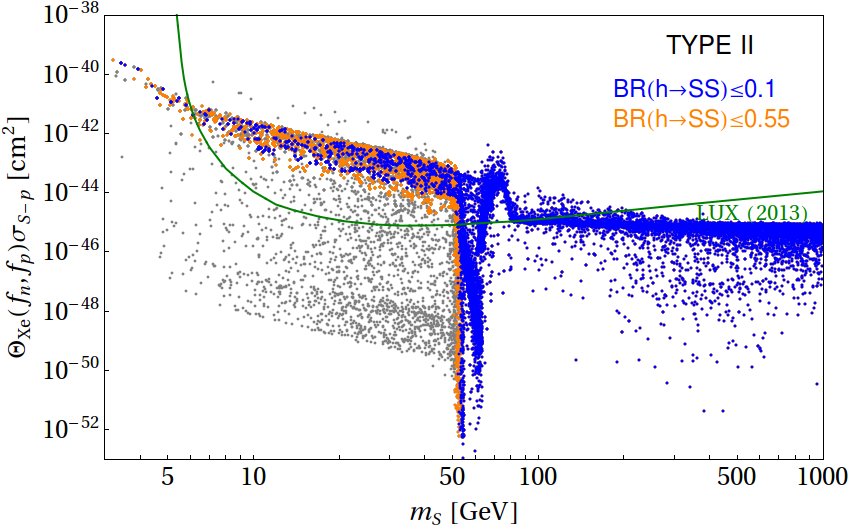}
\end{center}
\vspace{-.25in}
\caption{
Cross section for DM - proton scattering for the \typeii\ model rescaled by the function $\Theta_{\rm X}$ defined in \eq{DDtheta}, where X = Xe for a Xenon-based detector.
All points plotted satisfy  the preLUX constraints {\it except } $\br(h\to SS)\leq 0.1$ (\ie\ they satisfy the theoretical constraints for 2HDMS, 2HDM fitting at 68\% C.L. and the constraint on $\omghsq$). In the left-hand plot, for the light purple points the  ratio $f_{n}/f_{p}$ is within the range (0.95, 1.05).  For the darker purple points $-0.8 \leq f_{n}/f_{p} \leq -0.6$. The right-hand plot displays points that obey $\br(h\to SS) \leq 0.1$ in blue (\ie\ they obey the full set of preLUX constraints), while the orange points obey only the weaker limit of $\br(h\to SS)\leq 0.55$.
}
\label{CS_ms_all_typeII_a}
\end{figure}

In order to present the overall picture for \typeii, we adopt the parameters in~\cite{Feng:2011vu} to calculate the rescaling factor $\Theta_{\rm Xe}$ for the Xenon-based detectors and present the $\overline{\sigma}_{DM-p}$ cross sections in Fig.~\ref{CS_ms_all_typeII_a}. In the left plot, we impose all preLUX constraints (including $\omghsq$ in the $3\sigma$ window)  other than $\br(h\to SS)\leq 0.1$. Points with $f_n/f_p\sim 1$ (for which $\Theta_{\rm Xe} \sim 1$)  are singled out as are points with $f_n/f_p\sim -0.7$.  
Comparing with the right plot, one can find that only the former points can have  $\br(h\to SS)\leq 0.1$, and only a subset of these can obey the LUX limits.  Basically, we find that obtaining correct $\omghsq$ while at the same time having $\br(h\to SS)\leq 0.1$ (or even $\leq 0.55$) is not possible for the $f_n/f_p\sim -0.7$ points in the low-$m_S$ region. 

\begin{figure}[h]
\begin{center}
\includegraphics[width=0.45\textwidth]{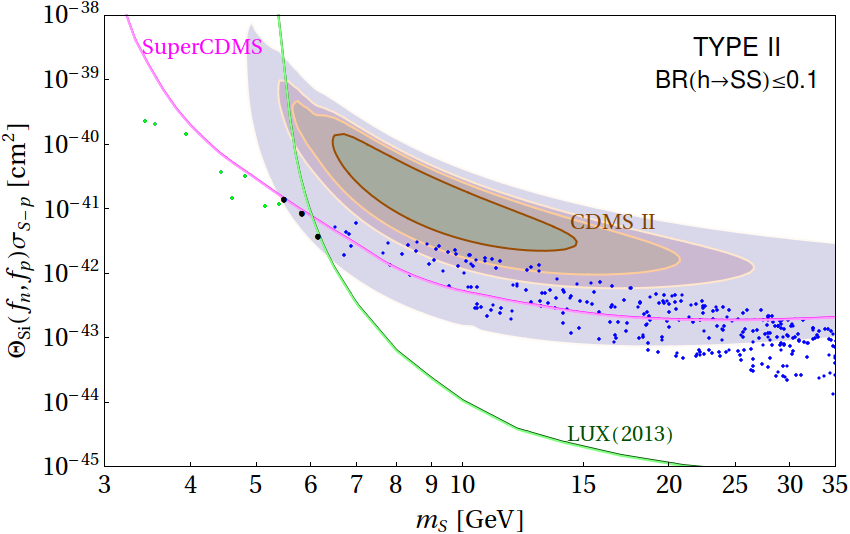}
\includegraphics[width=0.45\textwidth]{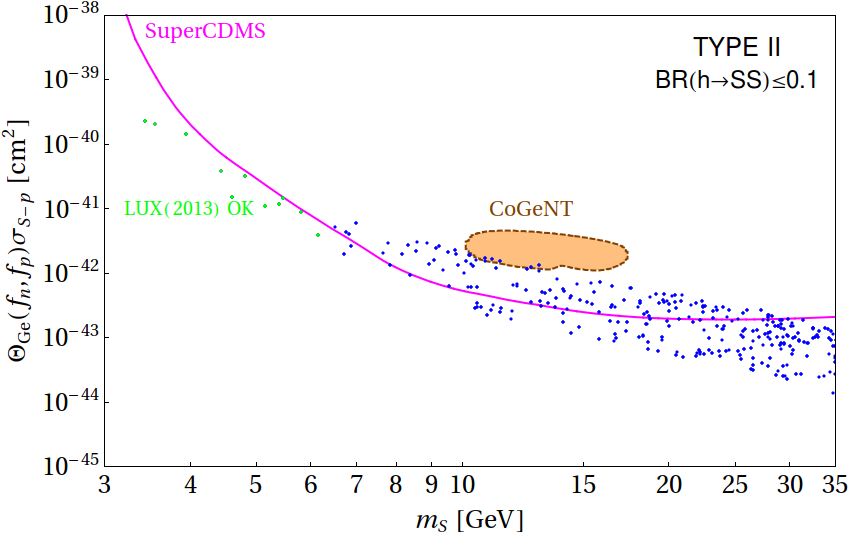}
\end{center}
\vspace{-.25in}
\caption{
Cross section for DM - proton scattering for the \typeii\ model rescaled by the function $\Theta_{\rm X}$ defined in \eq{DDtheta}, where X = Si for a Silicon detector (CDMS~II) on the left and X=Ge for the Germanium detector (CoGeNT) on the right. All points satisfy all the preLUX constraints (\ie\ they satisfy the theoretical constraints for 2HDMS, 2HDM fitting at 68\% C.L.,  $\br(h\to SS)\leq 0.1$ and the constraint on $\omghsq$).
 The CDMS~II contours shown are at  68\%, 90\%, 95\% and 99\% C.L.
 The CoGeNT contour is the 90\% C.L. level contour. Light green points are allowed by LUX results. 
The larger black points are those allowed by both SuperCDMS and LUX and that also lie within the 99\% C.L. CDMS~II contour.
The pink and light pink lines (almost degenerate) correspond to the SuperCDMS limit, after rescaling  from the SuperCDMS Germanium target to the CDMS-II Silicon target using $f_n/f_p=1.05$ and $1.25$ (the minimum and maximum values shown in Fig.~\ref{BRh_fnfp_all_typeII} for $\br(h\to SS)\leq 0.1$ when $\mS\leq 35\gev$).
Also shown by the dark green lines is the rescaled LUX limit, $\sigma^{\rm LUX}_{\rm Si} = \sigma^{\rm LUX} \Theta_{\rm Si}(f_n,f_p)/\Theta_{\rm Xe}(f_n,f_p)$, using the same two $f_n/f_p$ values.
}
\label{CS_ms_all_typeII_b}
\end{figure}

To explore in more detail the level of  inconsistency between the LUX and SuperCDMS limits and the positive signal regions for CDMS~II and CoGeNT, we present Fig.~\ref{CS_ms_all_typeII_b} which focuses on the $\mS\leq 35\gev$ mass range.  All plotted points obey the full set of preLUX constraints (including $\br(h\to SS)\leq 0.1$).  For the left figure, we have rescaled the  DM-proton scattering cross section predicted for a given point by the factor  $\Theta_X$, see  \eq{DDtheta}, as computed for X = Si in order to compare to the positive signal region found by the CDMS~II Silicon detector.  We also display the relevant limits from the SuperCDMS experiment.  These are $f_n/f_p$ dependent. The two lines correspond to the SuperCDMS limit after rescaling from the SuperCDMS Germanium target to the CDMS-II Silicon target. We rescaled $\sigma^{\rm SuperCDMS}_{\rm Si} = \sigma^{\rm SuperCDMS} \Theta_{\rm Si}(f_n,f_p)/\Theta_{\rm Ge}(f_n,f_p)$ using $f_n/f_p=1.05$ and $1.25$ --- the minimum and maximum values shown in Fig.~\ref{BRh_fnfp_all_typeII} for $\mS\leq 35\gev$ when $\br(h\to SS)\leq 0.1$ is imposed. We see that for the predicted range of $f_n/f_p$ the resulting rescaling is fairly minimal and those two limits are almost degenerate. Also shown by another two lines is the rescaled LUX limit, $\sigma^{\rm LUX}_{\rm Si} = \sigma^{\rm LUX} \Theta_{\rm Si}(f_n,f_p)/\Theta_{\rm Xe}(f_n,f_p)$, using the same two $f_n/f_p$ values.
From this plot, we observe that there are a few points (the large black points) with $\mS\sim 5.5-6.2\gev$ that lie below both the rescaled LUX  limits and rescaled SuperCDMS limits.  Further, although these points lie below the $2\sigma$ (95\% C.L.) contour of the positive signal region of CDMS~II, they do fall within the $3\sigma$ (99\% C.L.) contour.  Thus, the 2HDMS \typeii\ model  allows consistency between the  CDMS~II signal region (at 99\% C.L.) and the  SuperCDMS and LUX limits for a small range of low $m_S$.  

It is perhaps important to understand the points in Fig.~\ref{CS_ms_all_typeII_b} with low $\mS$ that obey LUX {\it and}  SuperCDMS constraints  in the case of the \typeii\ model. Their properties appear in Table~\ref{typeIIlowmSgood}. All have low $\tanb$, very modest $\mH$ with $m_A,\mhpm$ somewhat larger (in the $300-600\gev$ range).

For the right figure, we rescale  $\sigma_{{\rm DM}-p}$ using  $\Theta_X$ as computed for X=Ge in order to compare to the potential signal region for the CoGeNT Germanium detector.
 We find points consistent with all pre-LUX constraints within the CoGeNT 90\% C.L. signal region for $m_S\sim 10-15\gev$.
 However,  the entire CoGeNT signal region is excluded by the SuperCDMS limit (no relative rescaling required since both are for a Germanium target) and by the  LUX limit  as indicated by the point coloring (where these limits have been rescaled using the $f_n/f_p$ value for a given point to determine whether or not the point is excluded).
 
 In  the case of both the CDMS~II figure and the CoGeNT figure, 
we note that allowing  $\br(h\to SS)$ larger than $0.1$ does not allow points  much above those already shown, but  rather increases the density of points where points are already shown.

\begin{table}[htdp]
\caption{
Summary of the properties of the 2HDM Type II points in Fig.~\ref{CS_ms_all_typeII_b} which make it possible to realize $m_S<50\gev$, after imposing the full set of preLUX constraints together with the LUX and SuperCDMS bounds. All masses are given in GeV units. 
 }
\vspace*{-3mm}
\begin{center}
\begin{tabular}{|c|c|c|c|c|c|c|c|}
\hline
$\tanb$ & $\sina$ & $m_H$ & $m_A$  & $\mhpm$  &  $m_{12}^2$ & ($m_S$ [GeV], $\log{\Theta_{\rm Xe}(f_n,f_p) \sig_{S-p}} [{\rm cm}^2])$ \cr
\hline
2.092 & -0.41 & 138 & 451  & 399 & -12642 & (3.44, -39.65); (3.56,-39.69); (3.95, -39.85)\cr
\hline
3.121  & -0.282 & 187  & 546 &  571 &  8943 & (4.82, -40.50); (5.48, -40.83)\cr
\hline
2.192  & -0.394 & 209 & 488 & 503 & 7518  & (5.40, -40.93)\cr
\hline
1.728  & -0.476 & 177  & 318 &  389 &  9382 & (5.16, -40.97)\cr
\hline
1.789  & -0.461 & 198 &  420 &  430 & -6594  & (4.44,-40.43); (5.15, -40.96)\cr
\hline
1.488  & -0.528 & 157 &  553  & 576 & -10094  & (4.61, -40.83)\cr
\hline
2.375  & -0.363 & 259 & 260 & 339 &  15899 & (5.83, -41.05)\cr
\hline
\end{tabular}
\end{center}
\vspace*{-6mm}
\label{typeIIlowmSgood}
\end{table}%

As in the case of \typei, we could plot  $\lamh$ and $\lamH$ vs. $\mS$ for the \typeii\ points that obey preLUX constraints.  The resulting point distributions look very similar to those shown in Fig.~\ref{lamhH_ms_typei}.

\subsection{Summaries}
\label{sec:summaries}

It is perhaps useful to summarize what Type I and II models predict with regard to the invisible decays of the heavier $H$ and how this will impact possibilities for detecting the $H$ in upcoming LHC runs.
For $m_S\lsim 55\gev$, the $\br(h\to SS)\leq 0.1$ constraint required by a good $h$ fit to the $125.5\gev$ data  implies that $\lamh$ is small  and this indirectly impacts $\br(H\to SS)$. Before imposing the LUX limits, we find that  $BR(H\to SS)$ can have a number of semi-discrete values below 1, the discreteness being associated with particular  2HDM 68\% C.L.  points, but for the bulk of $m_S\lsim 55\gev$ points one has $\br(H\to SS)\gsim 0.9$. Of course, we have seen above that once the LUX and SuperCDMS limits are imposed all the low-$m_S$  points are eliminated in the \typei\ case, whereas in the \typeii\ case a handful of points survive in the $\mS\leq 6\gev$ region. Once $m_S\gsim 55\gev$, $\br(h\to SS)$ is automatically small or zero and constraints on $\lamh$ in the singlet sector scan are greatly relaxed.  As a result, $\br(H\to SS)$ can take on most any value for $m_S\lsim 200\gev$, declining to small values once $m_S\gsim 500\gev$.

As regards $H$ detection, we first note that since the $HVV$ couplings are small (since the $hVV$ coupling must be large for a good Higgs fit) the $Z+inv$ final state LHC data do not currently constrain $\br(H\to SS)$, and in future runs very high integrated luminosity would be needed to have any hope of seeing a signal in this channel.  Further, if $H\to SS$ decays  are dominant this would reduce the strength of the $H$ signals in other production/decay modes, such as $gg\to H\to \tau\tau$, and thus decrease the prospects for $H$ discovery as outlined in \cite{Dumont:2014wha}.  In such instances, experimental sensitivity to the $H$ may have to rely on $gg\to H$ production with a jet or photon tag of the invisible $H\to SS$ final state.  


We now turn to an expanded discussion of the summary given above in which we split the scalar mass $m_S$ into three regions, depending on the status of the exotic decay $h\to SS$: 
\begin{itemize}
\item low mass region ($1-55\gev$) where the decay is open and could be substantial without $\lamh$ being very small;
\item resonance region ($55-70\gev$) where $\mS$ is not far from the $h$ pole location.  For $\mS<\mh/2$, one finds that, after imposing P+S+U+EW, $\lamh$ is sufficiently limited that $\br(h\to SS)\leq 0.1$.  In fact, in this region, the strongest constraint on $\lamh$ comes from the need to avoid too much annihilation.
\item high mass region ($70-1000\gev$) where the decay is absolutely closed.
\end{itemize}
Note that we adopt different scan strategies in these regions of $m_S$ so as to achieve a maximum density around the most interesting points that pass all theoretical and experimental constraints.  The scans are also preformed in a different way for \typei\ and \typeii\ models.


\subsubsection{Low mass region}
\label{low_mass}

\vspace*{-2mm}
\begin{figure}[b]
\begin{center}
\includegraphics[height=0.34\textwidth ]{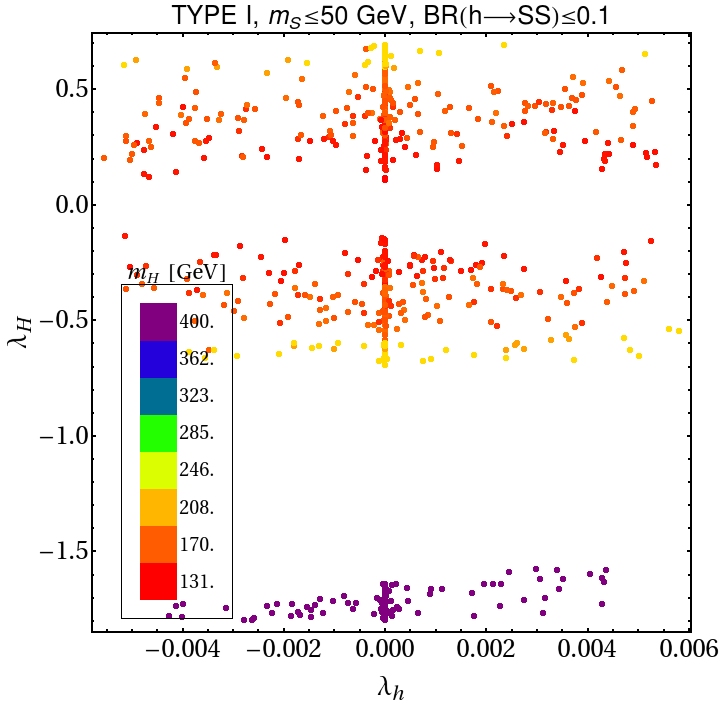}
\hspace{8mm}
\includegraphics[height=0.34\textwidth ]{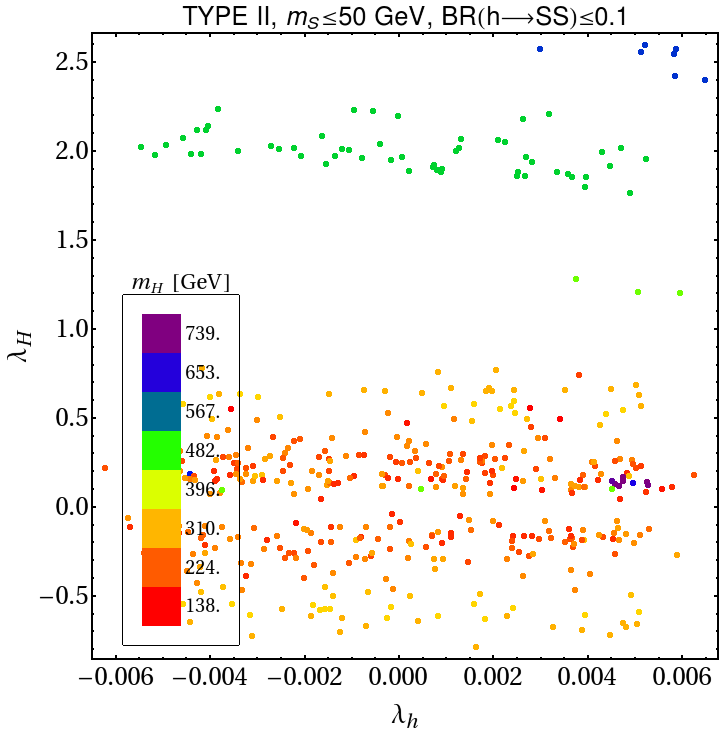}\\
\includegraphics[height=0.34\textwidth ]{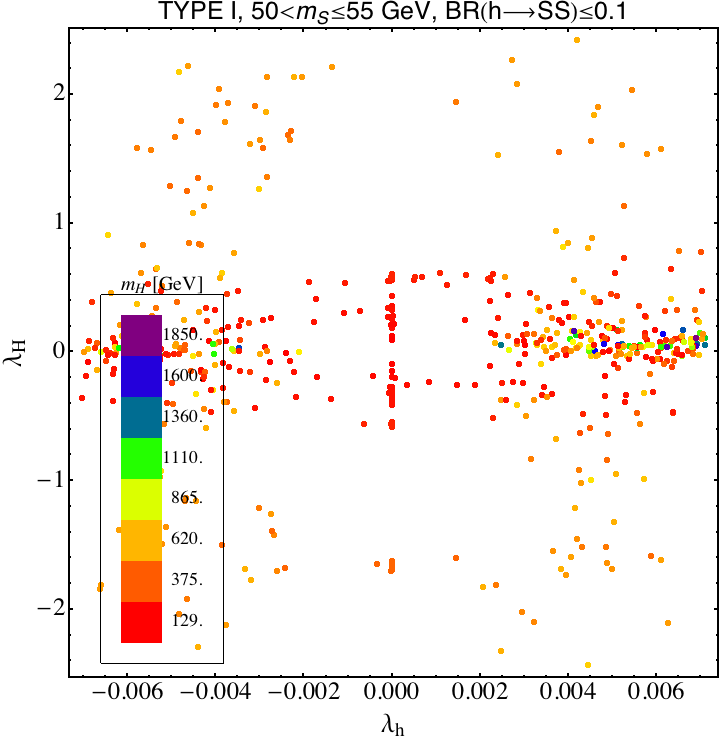}
\hspace{8mm}
\includegraphics[height=0.34\textwidth ]{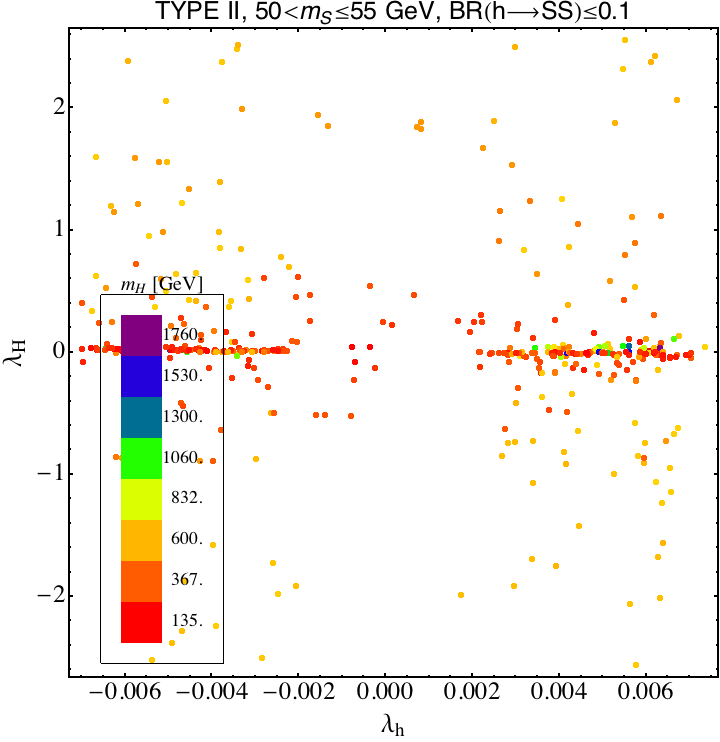}\\
\end{center}
\caption{Couplings $\lambda_h,\lambda_H$ allowed by the full set of preLUX constraints for the $\mS\leq 55\gev$ mass region. Points are temperatured according to $\mH$, with red points corresponding to the lowest $H$ mass for which a solution was found.  Upper figures are for $\mS\leq 50\gev$ while lower figures are for $50< \mS\leq 55\gev$. } 
\label{l1l2_low}
\end{figure}

As we have already noted, in the low mass region, the exotic decay $h\to SS$ could have a large branching ratio.  In the case where a singlet scalar is added to the pure SM,  one finds that the corresponding coupling of dark 
matter to the Higgs necessary to avoid overabundance of the relic $S$  is so large that   $\br(\hsm\to SS)\gsim 0.9$ \cite{He:2011gc,Drozd:2011aa}, thereby making a good fit of the $\hsm$ to the LHC Higgs data impossible.  
In the 2HDMS, one can keep $\br(h\to SS)$ small enough ($\leq 0.1$)   to avoid destroying the fit of the $h$ to the $125.5\gev$ Higgs data if  $\lambda_h \ll 1$.   Nonetheless, correct $\omghsq$ can be achieved because
in the 2HDMS the annihilation of DM is mediated not only by $h$ but also by $H$ (see Fig.~\ref{RelicDiag}). Therefore, the desired large cross section for $SS$ annihilation can be achieved  if $\lambda_H$ is sufficiently large when  $\lambda_h$ is small. This trend was already apparent in Fig.~\ref{lamhlamH}.  Here, we zero in on the $\mh\leq 55\gev$ region in  Fig.~\ref{l1l2_low}, where we have employed a special scan strategy designed to cover a large range of $f_n/f_p$ and small $\lamh$.  In the upper plots in Fig.~\ref{l1l2_low}, we require $\mS\leq 50\gev$ while the lower plots are for  $50 <\mS\leq 55\gev$.  In the latter case, we observe a hole in the vicinity of small $\lamh$ and $\lamH$ which expands to a gap in the former case due to the fact that points with $\mS\leq 50\gev$ are sufficiently far from the resonance region that $H$ exchange, \ie\ $\lamH\neq 0$, is needed for correct $\omghsq$.  In contrast, for points with $50<\mS\leq 55\gev$, for $\lamh\neq 0$ the $h$ alone can provide enough annihilation for correct $\omghsq$ even if $\lamH=0$.

As expected, the temperature plots show that, generally speaking,  the larger $\mH$ is the larger $\lamH$ must be for correct relic density (the $SS$ annihilation amplitude containing the ratio $\lamH/\mH^2$).  However, there is an exception in the case of the \typeii\ model; at large $\tanb$ ($\gsim 25$) one can have sufficient annihilation even if $\lamH/\mH^2$ is not large since the $Hb\anti b$ coupling is highly enhanced, $C^H_D\propto \tanb$, see \eq{ssxx}.  We observe a smattering of such points in the (upper) $\mS\leq 50\gev$ \typeii\ plot.  For these points,  the $SS \to b\bar{b}$ annihilation cross section is large enough to produce relic abundance within the experimental limit even though $|\lam_H| < 0.2$ and $m_H > 500 \gev$. 

\begin{figure}[h!]
\begin{center}
\includegraphics[height=0.25\textwidth]{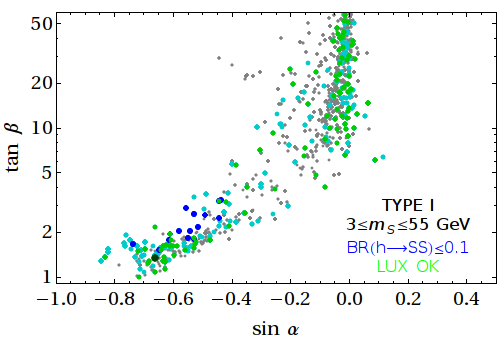}
\hspace{5mm}
\includegraphics[height=0.25\textwidth]{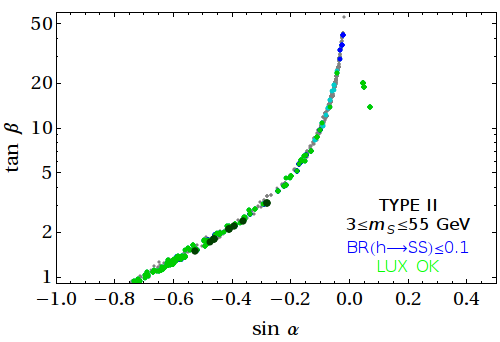}\\
\vspace{3mm}
\includegraphics[height=0.25\textwidth]{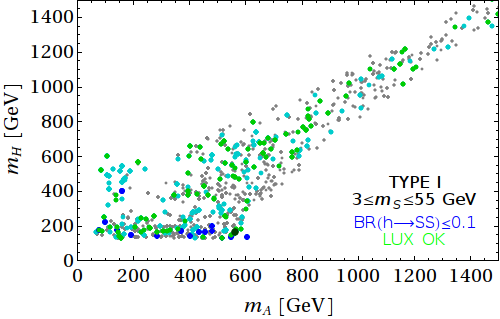}
\hspace{5mm}
\includegraphics[height=0.25\textwidth]{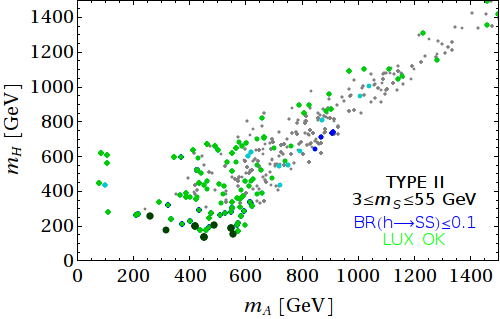}\\
\vspace{3mm}
\includegraphics[height=0.25\textwidth]{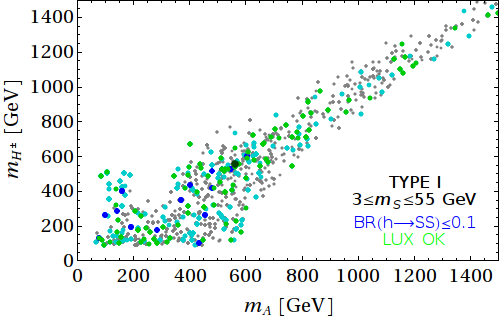}
\hspace{5mm}
\includegraphics[height=0.25\textwidth]{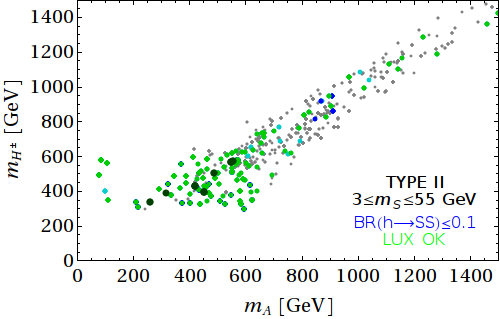}
\end{center}
\vspace*{-7mm}
\caption{We show how the full set of preLUX constraints on the singlet sector affects the 2HDM parameter space that we used for the singlet sector scans. We have required $\mS\leq 55\gev$.
Gray points satisfy all preLUX constraints aside from $\br(h\to SS)\leq 0.1$. Cyan and blue points satisfy in addition $\br(h\to SS)\leq 0.1$, \ie\ the full set of preLUX constraints. Cyan points have $50 < \mS\leq 55\gev$ while blue points have $\mS\leq 50\gev$.  The  green and dark green points satisfy the LUX bound as well as the full set of preLUX constraints, with dark green showing the $\mS\leq 50\gev$ points.}
\label{2hdm_par_low}
\end{figure}

We end this subsection with the plots of Fig.~\ref{2hdm_par_low} showing the regions of the 2HDM parameter space with $\mS\leq 55\gev$ that remain after imposing the full set of preLUX constraints.   The allowed regions are displayed in the $(\tanb,\sin\alpha)$, $(m_H,m_A)$ and  $(m_{H^\pm}, m_A)$ planes.  Different colors are used to distinguish  those points with $\mS\leq 50$ from those with $50< \mS \leq 55\gev$.  Also shown are those points that in addition satisfy the LUX limit.

\vspace*{-2mm}




\clearpage
\subsubsection{Resonance region}
\label{res_reg}

\begin{figure}[h]
\begin{center}
\includegraphics[height=0.34\textwidth ]{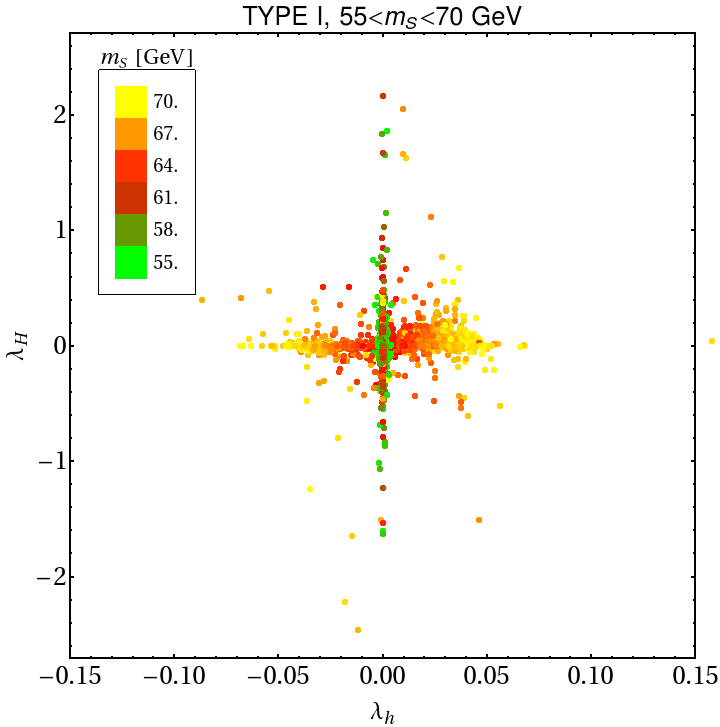}
\hspace{8mm}
\includegraphics[height=0.34\textwidth ]{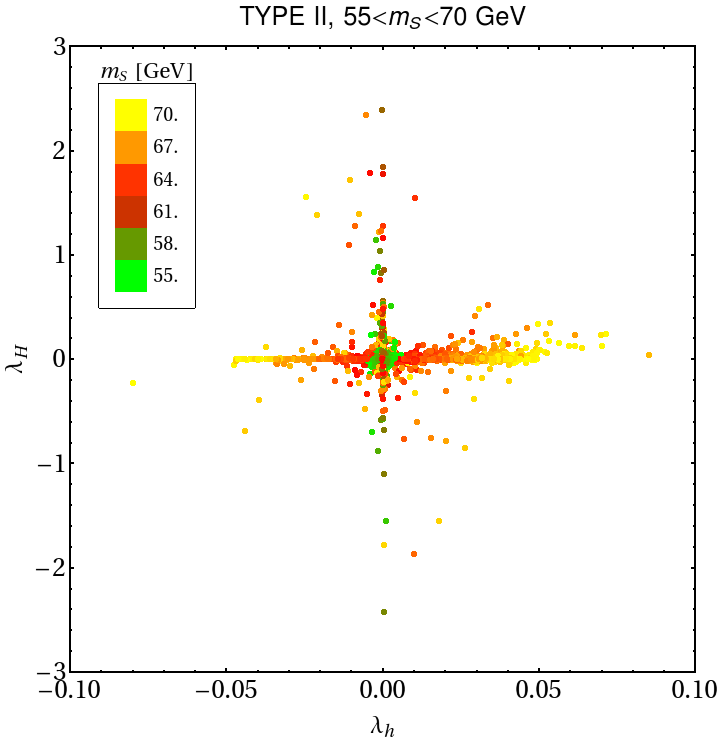}\\
\vspace{5mm}
\includegraphics[height=0.34\textwidth ]{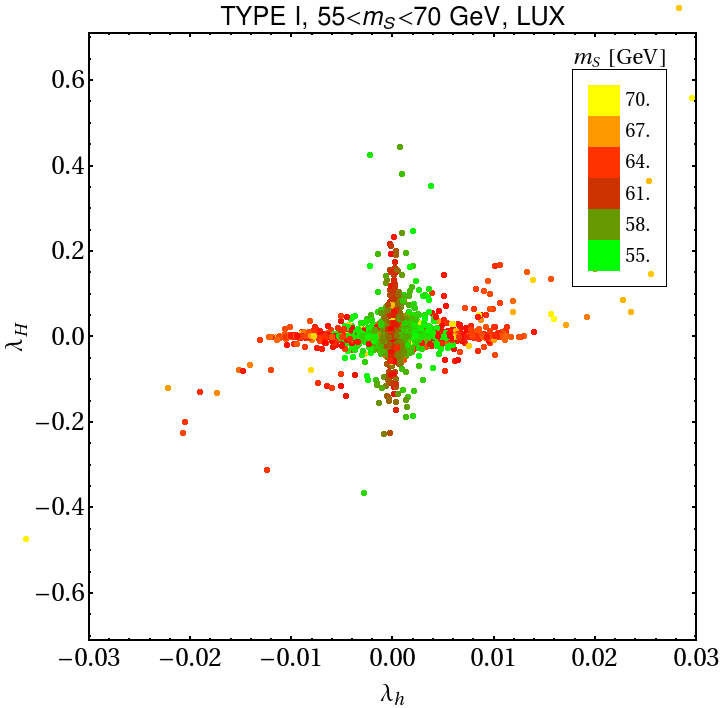}
\hspace{8mm}
\includegraphics[height=0.34\textwidth ]{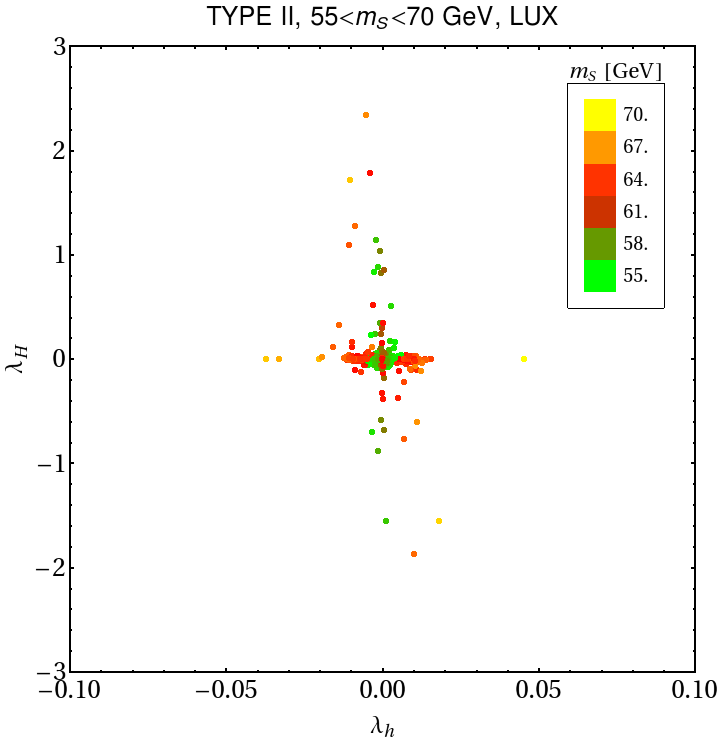}
\end{center}
\caption{In the upper plots we show the couplings ouplings $\lambda_h,\lambda_H$ allowed by the full set of preLUX constraints when $\mS$ is in  the resonance mass region, $55\gev < \mS \leq 70\gev$.  In the lower plots, we show the points of the upper plots that are also consistent with the 
LUX limit on DM scattering.}
\label{l1l2_res}
\end{figure}

In this subsection we focus on the $h$ resonance region, $55\gev <\mS\leq 70\gev$, which is defined such that the $h$ is near the pole of $SS$ annihilation, $m_S \approx m_h/2$. In this region, the annihilation of $SS$ into SM particles is mainly mediated through exchanging an $s$-channel $h$ (unless the $H$ is not much heavier than the $h$, $m_H \approx m_h$). For a given magnitude of $\lamh$ (and $\lamH$ when $\mH$ is close to $\mh$), the annihilation cross section is greatly enhanced  in the resonance region, as seen in Fig~\ref{CS_ms_all_typeI} for \typei\ and Fig.~\ref{CS_ms_all_typeII_b} for \typeii, respectively. In order to compensate for the resonance enhancement, $\lambda_h$ and/or $\lambda_H$ 
in the resonance region must be small in order to reproduce the observed DM abundance, as shown in Fig.~\ref{l1l2_res}.\footnote{
In this mass range we scan over $\lam_h$ and $\lam_H$ from $(10^{-4},10^{-1})$ and $(10^{-4},4\pi)$, respectively, with logarithmic density.} In the upper plots of this figure, we have imposed the full set of preLUX constraints including $BR(h\to SS)\leq 0.1$. In the lower plots, we have required that the LUX limits also be obeyed.  We observe that this latter requirement reduces further the magnitudes of $\lamh$ and $\lamH$.

 
We also note that in this resonance region $BR(H\to SS)$ is typically large, between $\sim 0.1$ and $\sim 0.9$.  If we were to repeat the plots of Fig.~\ref{2hdm_par_low} for this case, we would find little change in the regions allowed, just an increase in point density.  Indeed, very few of the starting 2HDM red points of Fig.~\ref{2hdmsample} are eliminated by the preLUX constraints, implying that the regions shown are nearly identical to those for the original 2HDM points sampled.
The reason for this is that once we are in the resonance region  correct relic density can almost always be obtained by judiciously choosing  $\lamh$ and $\lamH$. 
%

\subsubsection{High mass region}
\label{hig_mass}

\begin{figure}[h]
\begin{center}
\includegraphics[height=0.34\textwidth ]{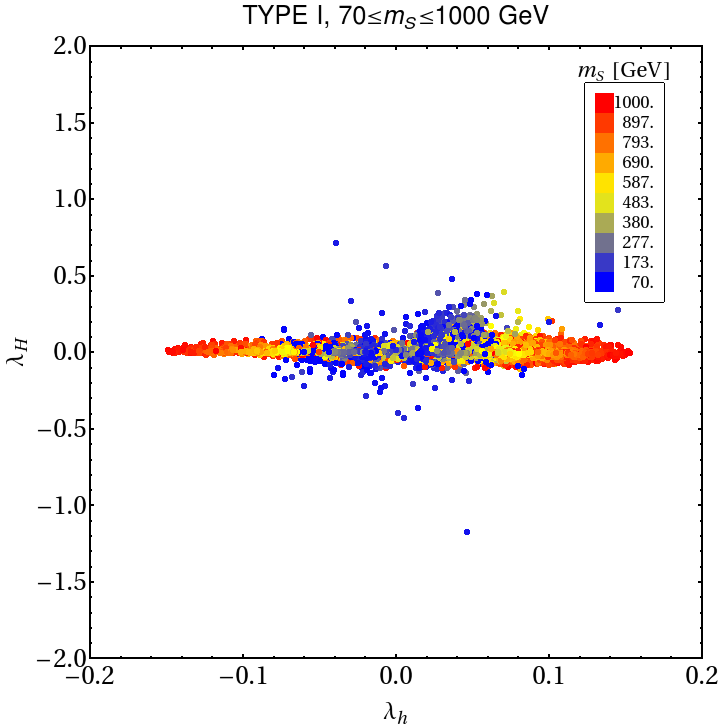}
\hspace{8mm}
\includegraphics[height=0.34\textwidth ]{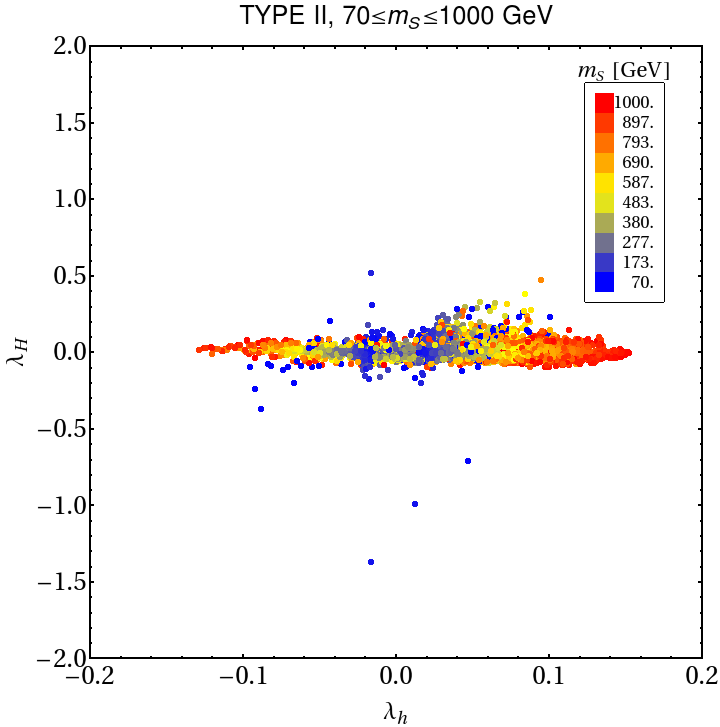}\\
\vspace{5mm}
\includegraphics[height=0.34\textwidth ]{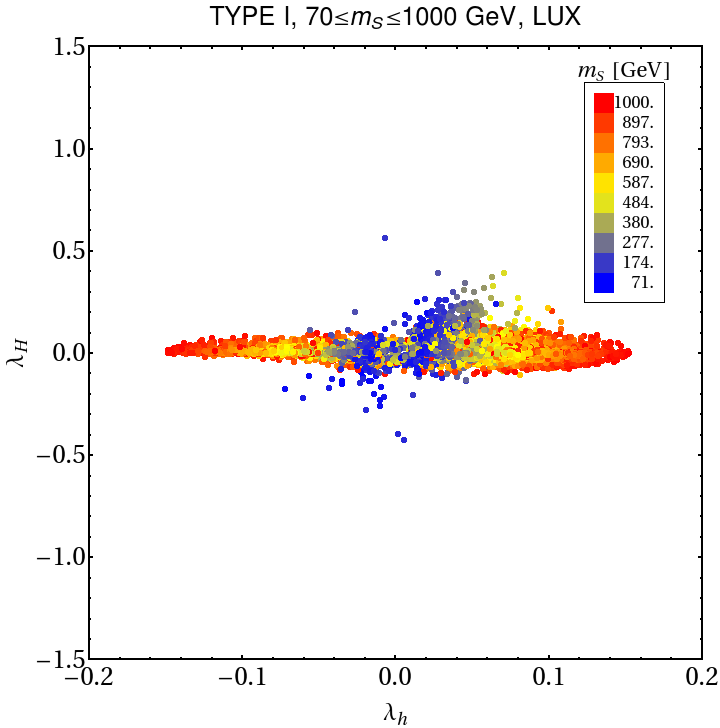}
\hspace{8mm}
\includegraphics[height=0.34\textwidth ]{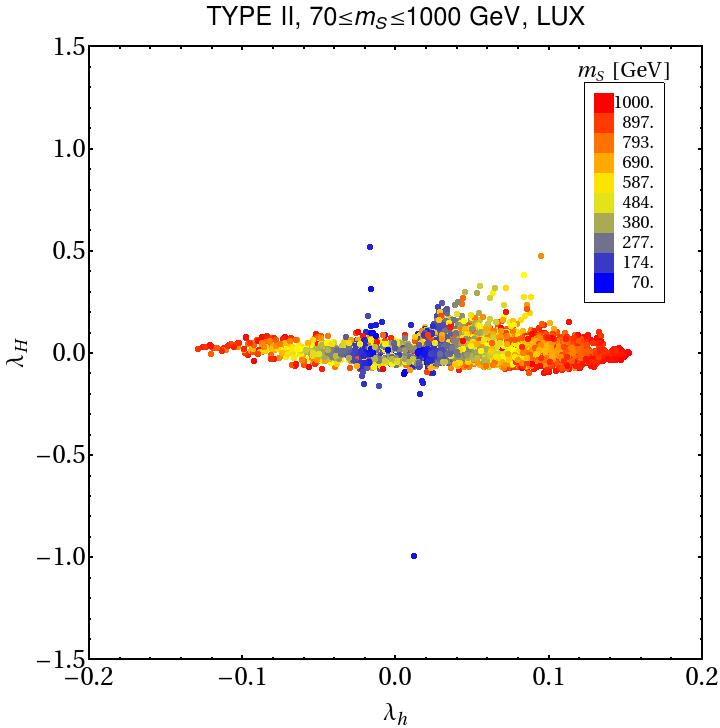}
\end{center}
\caption{Couplings $\lambda_h,\lambda_H$ allowed by the full set of preLUX constraints for $\mS$ within the high mass region. The LUX limit is imposed on the points shown in the lower plots.}
\label{l1l2_lar}
\end{figure}

The high mass region is defined as $70\leq m_S \leq 1000 \gev$. In our study, the parameters $\kap_1$ and $\kap_2$ in the extra singlet sector are both scanned over in the range $(10^{-2},4\pi)$ with logarithmic density. Points surviving the full set of preLUX constraints are shown in the  ($\lambda_h$, $\lambda_H$) plane in the upper plots of Fig.~\ref{l1l2_lar}. The lower plots show the points that also survive the LUX bound. There, one can observe that for high $m_S$ there is an ample parameter space surviving the preLUX constraints together with the LUX bound on the spin-independent cross section of DM direct detection.  

\clearpage
\section{Conclusions}
\label{sec:conclusions}
We have analyzed the 2HDMS models obtained by extending the \typei\ and \typeii\  two-Higgs-doublet models to include a scalar gauge-singlet dark matter candidate, denoted $S$ with mass $\mS$.   We have discussed various theoretical and experimental constraints on the 2HDMS and how these constrain the additional (beyond the 2HDM) three parameters of the 2HDMS, $\mS$ and the trilinear $hSS$ and $HSS$ couplings.  We begin with the 2HDM fits of \cite{Dumont:2014wha} for the case where it is the lighter $h$ that is identified with the $\sim 125.5\gev$ state, in particular employing the 2HDM parameter space points for which the  combined LHC/Tevatron signal strengths are fit within the 68\% C.L. 
We then study the constraints on the singlet parameter space based on cosmological data, most particularly the observed $\omghsq$ and the LUX and SuperCDMS limits on DM-nucleon scattering. If $\mS> 55\gev$, 2HDMS parameter choices for which the 2HDMS is completely consistent with all the above data are plentiful in both the \typei\ and \typeii\ models. For $\mS\leq 55\gev$, requiring $\br(h\to SS)\leq 0.1$ in order to avoid destroying the fit of the $h$ to the LHC data makes it impossible (almost impossible) in the \typei\ (\typeii) model to find parameter points that give correct $\omghsq$ and satisfy both the LUX and SuperCDMS limits. Nonetheless, it is interesting to note that if we do not impose the LUX and SuperCDMS limits, for both model types $\mS<50\gev$-points with $\br(h\to SS)\leq 0.1$  and correct $\omghsq$ fall within one or more of the  CDMS~II, CRESST-II or CoGeNT signal regions. 


An important issue in the 2HDMS context is whether or not there is a possibility of isospin violation, $f_n/f_p\neq 1$.  
In the case of the 2HDMS \typei\ model, $f_n/f_p\simeq 1$ is inevitable.  This,  implies that despite the fact that all points with correct $\omghsq$  and $\br(h\to SS)\leq 0.1$ have $\sigma_{S-p}$ values falling within one or more of the CDMS~II, CRESST-II or CoGeNT signal regions,  they are simply inconsistent with the LUX  and SuperCDMS limits.

In the case of the 2HDMS \typeii\ model, a significant isospin violation in DM-nucleon scattering is possible, even reaching the value of $f_n/f_p\sim -0.7$  that would allow consistency of the LUX limit with the CDMS~II signal region. However, at the low $\mS$ values corresponding to the signal region, we find that points with $f_n/f_p\sim -0.7$ either have an $hSS$ coupling that is too large for $\br(h\to SS)\leq 0.1$  or  too small  to give sufficient annihilation to achieve correct $\omghsq$.  (At low $\mS$, the $H$ exchange contribution to $SS$ annihilation is not sufficient, given upper bounds on the $HSS$ coupling coming from perturbativity and unitarity.) Therefore, even though isospin violation might be present, the level of $f_n/f_p\sim -0.7$ cannot be made consistent with all phenomenological requirements.  The SuperCDMS limit further constrains the picture.  For the $f_n/f_p$ values predicted by the 2HDMS once correct $\omghsq$ and $\br(h\to SS)\leq 0.1$ are imposed, the isospin violation is only a small effect in comparing the Germanium target SuperCDMS limit to the Silicon target CDMS~II result.  In the end, one does find a few $m_S\sim5.5-6.2$ GeV-points  that lie below both the SuperCDMS and LUX limits and, interestingly, also fall within (but are outside) the 99\%~C.L. (95\%~C.L.) CDMS~II signal region. As typical for $\mS\leq 50\gev$, these points are such that $\br(H\to SS)$ is large, implying that jet- and/or photon-tagging will be needed for $H$ detection.

\clearpage

\appendix
\clearpage
\section{Vacuum stability}
\label{vac_stab}

We are considering 2HDMS scalar potential with $\lam_6=\lam_7=0$, $\lam_5 \in R$ and $\kap_3=0$:
\beq
\begin{split}
V(H_1, H_2,S)&=\frac{\lam_1}{2}(H_1^{\dg} H_1)^2+\frac{\lam_2}{2}(H_2^{\dg} H_2)^2+\lam_3 (H_1^{\dg} H_1)(H_2^{\dg} H_2) 
+\lam_4 |H_1^{\dg} H_2|^2 
 \\
&\quad
+\l[ \frac{\lam_5}{2}(H_1^{\dg} H_2)^2 + h.c.\right] 
 +\frac{1}{4!}\lam_{S}S^4 +\kap_{1}S^2 (H_1^{\dg} H_1)+\kap_{2}S^2(H_2^{\dg} H_2)
\end{split}
\eeq

We will use the following reparametrization of the potential \cite{Kaffas:2006nt,Grzadkowski:2009bt}:
\beq
\begin{split}
& |H_1| = r \cos \theta \sin \phi, \hspace{1cm} |H_2| = r \sin \theta \sin \phi, \\
& H_1^\dagger H_2= |H_1||H_2| \rho e^{i\gamma}, \hspace{0.52cm} S= r \cos  \phi \\
\end{split}
\eeq

Because  $\frac{ H_1^\dagger}{|H_1|} \cdot \frac{H_2}{|H_2|}$ is a product of unit spinors, it is a complex number $\alpha+i\beta$ such that $|\alpha+i\beta|<=1$.  I can rewrite it in polar coordinates as $\alpha+i\beta = \rho e^{i \gamma}$ with $\rho \in (0,1)$. Rewriting the potential in terms of $r, \theta, \phi, \rho, \gamma$ we get
\beq
\begin{split}
V/r^4&=\left(
\frac{\lam_1}{2}\cos^4 \theta+\frac{\lam_2}{2}\sin^4\theta+\lam_3 \cos^2\theta \sin^2\theta +\lam_4 \rho^2 \sin^2\theta \cos^2\theta  +\lam_5 \rho^2 \sin^2\theta \cos^2\theta\cos(2\gamma) \right)\sin^4\phi
 \\
&\quad
 +\frac{1}{4!}\lam_{S}\cos^4\phi +\kap_{1}\cos^\theta \sin^\phi \cos^2\phi+\kap_{2} \sin^2\theta \cos^2\phi \sin^2\phi
\label{potangles}
\end{split}
\eeq
where $V = V(\cos^2 \theta, \sin ^2 \phi, \cos(2\gamma), \rho)$ and this parameters change in the following ranges:
\beq
\begin{split}
& x= \cos^2 \theta  \in (0,1), \hspace{1.5cm}  y= \sin ^2 \phi   \in (0,1), \\
& z = \cos(2\gamma)  \in (-1,1), \hspace{1cm}  \rho \in (0,1) 
\end{split}
\eeq
and we can rewrite the potential in the final form:
\beq
\begin{split}
V/r^4&=\left(
\frac{\lam_1}{2}x^2+\frac{\lam_2}{2}(1-x)^2+\lam_3 x (1-x) +\lam_4 \rho^2 x (1-x) +\lam_5 \rho^2 x (1-x) z \right)y^2
 \\
&\quad
 +\frac{1}{4!}\lam_{S}(1-y)^2 +\left(\kap_{1}x+\kap_{2}(1-x) \right) y(1-y)
\label{potxyz}
\end{split}
\eeq

To assure stability potential has to be bounded from below, which means that in the limit of infinite fields $V$ has to approach $+\infty$. Therefore the necessary and sufficient consition for tree level stability of our theory is a positive minimum of $V/r^4$ in (\ref{potxyz}). Let's use a following lemma to find constraints on potential parameters for which the minimum is indeed positive:

\hspace{6cm} \textit{Lemma 1}
\beq
\begin{split}
& f(\xi)= a \xi^2 + b(1-\xi)^2 + c \xi (1-\xi) >0, \,\,\,\, \xi \in (0,1) \iff
a>0, b>0, c>-\sqrt{4ab}
\end{split}
\eeq

This can be easily shown to be true using basic properties of a quadratic function.  
\textit{Lemma 1} for $\xi=y$ leads to the following set of constraints:
\beq
\begin{split}
& A = \left(
\frac{\lam_1}{2}x^2+\frac{\lam_2}{2}(1-x)^2+\lam_3 x (1-x) +\lam_4 \rho^2 x (1-x) +\lam_5 \rho^2 x (1-x) z \right) > 0\\
& B = \frac{1}{4!}\lam_{S} > 0\\
& C = \left( \kap_{1}x+\kap_{2}(1-x) \right) > -\sqrt{4 A\,B}
\end{split}
\eeq

where $A>0$ is the regular 2HDM constraint and  $B > 0$ leads to $\lam_{S} > 0$. We can rewrite $C > -\sqrt{4 A\,B}$ in a way to use $Lemma \,I$ again.
\beq
0< 
\left(\frac{1}{12}\lam_{S}\lam_1 - \kap_{1}^2 \right) x^2
+\left(\frac{1}{12}\lam_{S}\lam_2-\kap_{2}^2\right)(1-x)^2 +
\left(- 2\kap_1\kap_2  +
\frac{1}{6}\lam_{S}(\lam_3 +\lam_4 \rho^2 +\lam_5 \rho^2 z )\right)x (1-x)
\eeq
This form leads to a following set of inequalities:
\beq
\begin{split}
& A' = \left(\frac{1}{12}\lam_{S}\lam_1 - \kap_{1}^2 \right) > 0,
\, B' = \left(\frac{1}{12}\lam_{S}\lam_2-\kap_{2}^2\right) > 0,
\, C' =\left(\frac{1}{6}\lam_{S}(\lam_3 +\lam_4 \rho^2 +\lam_5 \rho^2 z) \right) > -\sqrt{4 A'\,B'}
\end{split}
\eeq

The full set of stability constraints on 2HDMS potential is the following:
\begin{itemize}
\item $\lam_1, \lam_2, \lam_S >0$ 
\item  $\lam_3+\lam_4-|\lam_5|>-\sqrt{\lam_1 \lam_2}$
\item $\lam_3>-\sqrt{\lam_1 \lam_2}$
\item $\kap_1 > - \sqrt{\frac{1}{12}\lam_{S}\lam_1}$ 
\item $\kap_2 > - \sqrt{\frac{1}{12}\lam_{S}\lam_2}$ 
\item if  $\kap_1$ or $\kap_2<0$ then we have to satisfy also:
\begin{itemize}
\item [*] $- 2\kap_1\kap_2+\frac{1}{6}\lam_{S}\lam_3>-\sqrt{ 4\left(\frac{1}{12}\lam_{S}\lam_1 - \kap_{1}^2 \right)\left(\frac{1}{12}\lam_{S}\lam_2-\kap_{2}^2\right)}$
\item[*] $- 2\kap_1\kap_2+\frac{1}{6}\lam_{S}(\lam_3 +\lam_4-|\lam_5| )>-\sqrt{ 4\left(\frac{1}{12}\lam_{S}\lam_1 - \kap_{1}^2 \right)\left(\frac{1}{12}\lam_{S}\lam_2-\kap_{2}^2\right)}$\\
\end{itemize}
\end{itemize}

\section{Tree-level Perturbative Unitarity}
\label{per_unit}

A natural way for derivation of the tree-level unitarity constraints is to construct the full multi-state scattering matrix for all the physical scalar-scalar states (including possible states associated with longitudinal vector bosons) in the tree approximation at high enough energy (where the scattering is dominantly mediated by the direct quartic interactions), requiring its largest eigenvalue to be less than the upper limit, i.e. $16 \pi$. This way involves a large dimensional matrix for which determining the eigenvalues is usually impractical. Alternatively, this limitations for eigenvalues of scattering matrix can be obtained in any basis related to the physical basis by a unitarity transformation. It was shown in~\cite{Akeroyd:2000wc,Arhrib:2000is,Horejsi:2005da} that the derivation for the 2HDM is considerably simple in the basis of non-physical electroweak eigenstates although it still needs explicit work with components of Higgs doublets. An even simpler approach dealing with the initial doublets $H_1$ and $H_2$ was developed in~\cite{Ginzburg:2005dt}. In this paper we will adopt the technique introduced in~\cite{Arhrib:2000is} and extend the derivation to the 2HDM plus a singlet model, taking into account an additional singlet in constructing the full scattering matrix.

Following the previous studies~\cite{Akeroyd:2000wc,Arhrib:2000is,Horejsi:2005da}, we start with the most general 2HDMS potential in the HHG parametrization that is subject to 
\beq
\begin{split}
 V(\Phi_{1}, \Phi_{2}) = & \, \Lambda_{1} ( |H_{1}|^2- {1 \over 2} v_{1}^2)^2
+\Lambda_{2} (|H_{2}|^2-{1 \over 2} v_{2}^2)^2+
\Lambda_{3}[(|H_{1}|^2-{1 \over 2} v_{1}^2)+(|H_{2}|^2-{1 \over 2} v_{2}^2)]^2 \\
&  + \Lambda_{4}(|H_{1}|^2 |H_{2}|^2 - |H_{1}^+H_{2}|^2  )+
\Lambda_{5} [\Re(H^+_{1}H_{2})-{1 \over 2} v_{1}v_{2}]^2+ \Lambda_{6} [\Im(H^+_{1}H_{2})]^2\\
&  +\frac{1}{2} m_0^2 S^2 +\frac{1}{4!}\lam_{S}S^4 +\kap_{1}S^2 (H_1^{\dg} H_1)+\kap_{2}S^2(H_2^{\dg} H_2)
\label{higgspot}
\end{split}
\eeq
where the $\Lambda_i$ are real parameters. 

\beq
H_1=
\begin{pmatrix}
      w_1^+    \\
      \sqrthalf (v_1+h_1+i z_1)
\end{pmatrix},
\hspace{5mm}
H_2=
\begin{pmatrix}
      w_2^+    \\
      \sqrthalf (v_2+h_2+i z_2)
\end{pmatrix}
\eeq

In principle, one needs to examine the full scattering matrix, including all two-particle states made of the scalars (including the unphysical Goldstone bosons) $w^\pm_i$, $h_i$, $z_i$ ($i=1,2$) and $s$. Nonetheless, charge conservation and CP-invariance forbids some scattering processes, making it composed by four sub-matrices, as illustrated in~\cite{Arhrib:2000is,Horejsi:2005da}.

Without the help of the $s$ state originated from the singlet $S$ sector, one can construct the extra charged states ($w^+_1 s, w^+_2 s$) which decouple with the other charged states ($w^+_1 h_1, w^+_1 h_2, w^+_1 z_1, w^+_1 z_2, w^+_2 h_1, w^+_2 h_2, w^+_2 z_1, w^+_2 z_2$)  existed in the 2HDM. Only the last two terms in eq.(\ref{higgspot}) handle the scattering matrix under this basis, which is given by
\beq
\mathcal{M}=\begin{pmatrix}
2 \kap_1 &  0 \\
0 & 2 \kap_2
\end{pmatrix}
\eeq
It is apparently diagonal and has eigenvalues $2\kap_1,2\kap_2$.

However, the neutral states under the basis ($w^+_1 w^-_1, w^+_2 w^-_2, {z_1 z_1 \over \sqrt{2}}, {h_1 h_1\over \sqrt{2}}, {z_2 z_2\over \sqrt{2}}, {h_2 h_2\over \sqrt{2}}, {ss\over \sqrt{2}}$) is more involved. We have to write out the full scattering matrix,
\beq
\mathcal{M}=\begin{pmatrix}
   4\Lam^+_{13}  &  2\Lam_3+{\Lam^+_{56} \over 2}  & \sqrt{2}\Lam^+_{13}  & \sqrt{2}\Lam^+_{13}  & \sqrt{2} \tilde{\Lam}^+_{34}  & \sqrt{2} \tilde{\Lam}^+_{34}  & \sqrt{2} \kap_1\\
   2\Lam_3+{\Lam^+_{56} \over 2}  &   4\Lam^+_{23}  & \sqrt{2} \tilde{\Lam}^+_{34}  &  \sqrt{2} \tilde{\Lam}^+_{34}  & \sqrt{2}\Lam^+_{23}  & \sqrt{2}\Lam^+_{23}  & \sqrt{2} \kap_2\\
   \sqrt{2}\Lam^+_{13}  & \sqrt{2} \tilde{\Lam}^+_{34}  & 3\Lam^+_{13}  & \Lam^+_{13}  & \tilde{\Lam}^+_{35}  & \tilde{\Lam}^+_{36}  &  \kap_1\\
   \sqrt{2}\Lam^+_{13}  & \sqrt{2} \tilde{\Lam}^+_{34}  & \Lam^+_{13}  & 3\Lam^+_{13}  & \tilde{\Lam}^+_{36}  & \tilde{\Lam}^+_{35}  &  \kap_1\\
   \sqrt{2} \tilde{\Lam}^+_{34}  & \sqrt{2}\Lam_{23}  & \tilde{\Lam}^+_{35}  &  \tilde{\Lam}^+_{36} & 3\Lam_{23}  & \Lam_{23}  & \kap_2\\  
   \sqrt{2} \tilde{\Lam}^+_{34}  & \sqrt{2}\Lam_{23}  & \tilde{\Lam}^+_{36}  &  \tilde{\Lam}^+_{35} & \Lam_{23}  & 3\Lam_{23}  & \kap_2\\   
   \sqrt{2} \kap_1  & \sqrt{2} \kap_2  & \kap_1  & \kap_1  & \kap_2  & \kap_2  &  {1 \over 2}\lam_S
\end{pmatrix}
\eeq
where $\Lam^+_{ij} = \Lam_i+\Lam_j$ and $\tilde{\Lam}^+_{ij} = \Lam_i+{\Lam_j \over 2}$ with 
\bea \label{convertback}
&&\Lam_1=\half \left[\lam_1-(\lam_3+\lam_4+\lam_5)+2m_{12}^2/(v^2\sb\cb)\right]\,,\nonumber\\
&&\Lam_2=\half \left[\lam_2-(\lam_3+\lam_4+\lam_5)+2m_{12}^2/(v^2\sb\cb)\right]\,,\nonumber\\
&&\Lam_3=\half \left[(\lam_3+\lam_4+\lam_5)-2m_{12}^2/(v^2\sb\cb)\right]\,,\nonumber\\
&&\Lam_4=2m_{12}^2/(v^2\sb\cb)-\lam_4-\lam_5\,,\nonumber\\[5pt]
&&\Lam_5=2m_{12}^2/(v^2\sb\cb)\,,\nonumber\\[5pt]
&&\Lam_6=2m_{12}^2/(v^2\sb\cb)-2\lam_5.\nonumber
\eea

The analytical form of eigenvalues are
\bea
b_{\pm}&=&\Lam_1+\Lam_2+2\Lam_3 \pm \sqrt{(\Lam_1-\Lam_2)^2+{1\over 4}(2\Lam_4-\Lam_5-\Lam_6)^2}\\
c_{\pm}&=&\Lam_1+\Lam_2+2\Lam_3 \pm \sqrt{(\Lam_1-\Lam_2)^2+{1\over 4}(\Lam_5-\Lam_6)^2}
\eea
and the rest three ones $a_{1,2,3}$ comes from the cubic polynomial equation 
\beq
\begin{split}
& x^3 -12 \left(\tilde{\Lam}^+_{321} +{1\over 24}\lam_S \right)x^2 + \left[36\overline{\Lam}_{123}+(2\Lam_3-\tilde{\Lam}^+_{456})(10\Lam_3+\tilde{\Lam}^+_{456})+6\lam_S \tilde{\Lam}^+_{321}-4(\kap_1^2+\kap_2^2)\right] x \\
&-18\lam_S\overline{\Lam}_{123}-{1\over 2}\lam_S(2\Lam_3-\tilde{\Lam}^+_{456})(10\Lam_3+\tilde{\Lam}^+_{456})+24(\kap_1^2\Lam^+_{23}+\kap_2^2\Lam^+_{13})-8\kap_1\kap_2(4\Lam_3+\tilde{\Lam}^+_{456}) =0 \nonumber
\end{split}
\eeq
where $\tilde{\Lam}^+_{ijk} = \Lam_i+{\Lam_j + \Lam_k \over 2}$ and $\overline{\Lam}_{ijk}=\Lam_i \Lam_j+\Lam_i \Lam_k+ \Lam_j \Lam_k$.

Another two scattering submatrices under the neutral states basis ($h_1 s, z_1 s$) and ($h_2 s, z_2 s$) are both diagonal and has eigenvalues $2\kap_1$ and $2\kap_2$, respectively. 

Finally, the remaining $S$-matrix for $2-2$ process is unchanged and was already studied in \cite{Horejsi:2005da}. The eigenvalues which have significant impacts on preventing the unitarity bound are
\begin{equation}
  \begin{aligned}
   e_1 &= 2 \Lam_3 - \Lam_4 - \frac12 \Lam_5 + \frac52 \Lam_6 \\
   f_+ &= 2 \Lam_3 - \Lam_4 + \frac52 \Lam_5 - \frac12 \Lam_6 \\
   p_1 & = 2 (\Lam_3 + \Lam_4) - \frac 12 \Lam_5 - \frac 12 \Lam_6
  \end{aligned}
\end{equation}

Putting all together, it is thus sufficient to taking the inequalities into account as follows
\begin{equation}
 max\{|a_{1,2,3}|\}, |f_+|, |e_1|, |p_1|, 2\kap_1, 2\kap_2 \le 8\pi
 \label{uniq}
\end{equation}

\section*{Acknowledgments} 
JFG and YJ are supported in part by US DOE grant DE-SC-000999. YJ is also supported by LHC-TI fellowship US NSF grant PHY-0969510. AD and BG are supported by the National Science Centre (Poland) under research project, decision no DEC-2011/01/B/ST2/00438 and  by the Foundation for Polish Science International PhD Projects Programme co-financed by the EU European Regional Development Fund. JFG and YJ thank Warsaw University for support at various stage of this project. YJ also acknowledges partial support by the Mainz Dark Matter Workshop where the early plan of this project was presented. JFG and YJ thank Beranger Dumont and Sabine Kraml for their collaboration on the 2HDM data base employed as part of the present work.

\bibliographystyle{JHEP}
\bibliography{2hdms.bib}

\end{document}

\begin{figure}[h]
\begin{center}
\includegraphics[width=0.4\textwidth]{scanplots/typeI_largemS_allpoints/l1_ms_mh2.png}
\includegraphics[width=0.4\textwidth]{scanplots/typeII_largemS_allpoints/l1_ms_mh2.png}
\includegraphics[width=0.4\textwidth]{scanplots/typeI_largemS_allpoints/l2_ms_mh2.png}
\includegraphics[width=0.4\textwidth]{scanplots/typeII_largemS_allpoints/l2_ms_mh2.png}
\end{center}
\vspace{-.25in}
\caption{}
\label{l1l2_ms_high}
\end{figure}

\begin{figure}[h]
\begin{center}
\includegraphics[width=0.4\textwidth]{scanplots/typeI_largemS_allpoints/k1_ms_mh2.png}
\includegraphics[width=0.4\textwidth]{scanplots/typeII_largemS_allpoints/k1_ms_mh2.png}
\includegraphics[width=0.4\textwidth]{scanplots/typeI_largemS_allpoints/k2_ms_mh2.png}
\includegraphics[width=0.4\textwidth]{scanplots/typeII_largemS_allpoints/k2_ms_mh2.png}
\end{center}
\vspace{-.25in}
\caption{??? - $m_S$ from 63 GeV. Stability, unitarity, ewsb ok. $\Omega \pm 3\sigma$. Green points - LUX ok.}
\label{k1k2_ms_mH_high}
\end{figure}